\documentclass[a4paper,fleqn]{cas-sc}

\usepackage[numbers]{natbib}
\usepackage{amssymb}
\usepackage{amsmath}
\usepackage{graphicx}
\usepackage{subcaption}
\usepackage{multirow}
\usepackage{xcolor}
\usepackage{pifont}
\usepackage{diagbox}

\newcommand{\new}[0]{\textcolor{black}} 

\newcommand{\newr}[0]{\textcolor{black}} 

\begin{document}

\shorttitle{MXene-Based Metasurfaces Spectral Prediction Optimization}

\title[mode=title]{Optimizing Spectral Prediction in MXene-Based Metasurfaces Through Multi-Channel Spectral Refinement and Savitzky-Golay Smoothing}

\author[1,2]{Shujaat Khan}\corref{cor1}
\cortext[cor1]{Corresponding author}
\ead{shujaat.khan@kfupm.edu.sa}

\author[3]{Waleed Iqbal Waseer}
\ead{waleedwaseer@uestc.edu.cn}

\author[2]{Muhammad Shahid Jabbar}
\ead{muhammad.jabbar@kfupm.edu.sa}

\affiliation[1]{organization={Department of Computer Engineering, College of Computing and Mathematics, King Fahd University of Petroleum \& Minerals},
    city={Dhahran},
    postcode={31261},
    country={Saudi Arabia}}

\affiliation[2]{organization={SDAIA-KFUPM Joint Research Center for Artificial Intelligence, King Fahd University of Petroleum \& Minerals},
    city={Dhahran},
    postcode={31261},
    country={Saudi Arabia}}

\affiliation[3]{organization={School of Physics and State Key Laboratory of Electronics Thin Films and Integrated Devices, University of Electronics and Science and Technology of China},
    city={Chengdu},
    postcode={610054},
    country={China}}

\begin{abstract}
The prediction of electromagnetic spectra for MXene-based solar absorbers, where MXenes are a family of two-dimensional transition metal carbides and nitrides, is a computationally intensive task traditionally addressed using full-wave solvers. This study introduces an efficient deep learning framework incorporating transfer learning, multi-channel spectral refinement, and Savitzky-Golay smoothing to accelerate and enhance spectral prediction accuracy. The proposed architecture leverages a pretrained MobileNet version 2 model, fine-tuned to predict 102-point absorption spectra from ($64\times64$) metasurface designs. Additionally, the multi-channel spectral refinement module processes the feature map through multiple convolutional channels, enhancing feature extraction, while Savitzky-Golay smoothing mitigates high-frequency noise. Experimental evaluations demonstrate that the proposed model significantly outperforms baseline convolutional neural network and deformable convolutional neural network models, achieving an average root mean squared error of 0.0227, coefficient of determination ($R^2$) of 0.9563, and peak signal-to-noise ratio of 33.10 decibels. The proposed framework presents a scalable and computationally efficient alternative to conventional solvers, positioning it as a viable candidate for rapid spectral prediction in nanophotonic design workflows.
\end{abstract}


\maketitle

\section{Introduction}

Nanophotonic structures, including photonic crystals, plasmonic nanostructures, and engineered metasurfaces, have demonstrated exceptional capabilities in controlling electromagnetic wave propagation across various spectral ranges~\cite{tang2022topological, li2021recentplasmonic, cui2023guest}. The ability of structurally complex nanophotonic systems to offer optical functionalities that outperform those of natural materials has attracted increasing research interest. Key examples include flat lenses \cite{khorasaninejad2017metalenses} optical vortices \cite{huang2017volumetric}, plasmon-induced transparency and improved optical imaging \cite{naveed2021optical}, Bessel beam generation \cite{mehmood2016visible},  and perfect absorbers \cite{landy2008perfect}. In addition to these developments,  considerable research has focused on engineering a wide variety of absorber geometries for applications in sensing, filtering, and other photonic technologies.  However, Metal-insulator-metal (MIM) absorbers, in particular, have gained prominence as efficient platforms for absorption-based applications, such as solar harvesting, thermal emission, and photodetection. These structures exploit the interaction between dielectric and plasmonic resonances, enabling precise spectral tuning through tailored geometric designs~\cite{soni2023machine}.

Traditional absorber structures, including plasmonic and semiconductor-based materials, carbon nanotube arrays, and metal–dielectric multilayer films, face challenges by complex integration, high thickness requirements, and complex fabrication procedures. \cite{ding2023machine, mizuno2009black, chen2022plasmonic}.  MXene is two dimensional material represented by the general formula \( M_{n+1}X_nTx \), consisting of transition metal carbides or nitrides with tunable surface terminations, enabling precise control over electromagnetic properties. MXenes based absorbers have garnered considerable attention due to their superior electrical conductivity, broadband absorption characteristics, and ease of integration.
Recent studies have demonstrated MXene-based absorbers with enhanced solar absorption and thermal performance compared to conventional materials like carbon nanotubes~\cite{mizuno2009black, chen2022plasmonic}. Despite their advantages, designing MXene-based absorbers for specific spectral responses remains computationally demanding when using conventional electromagnetic solvers such as Finite-Difference Time Domain (FDTD), Finite Element Method (FEM), or Rigorous Coupled-Wave Analysis (RCWA)~\cite{yao2012highly, moharam1995stable}.

Recent advancements in deep learning have presented promising alternatives for predictive modeling of spectrums~\cite{liu2018training, park2021sersnet, khan2021switchable, park2023self}. \new{Convolutional neural networks (CNNs), including architectures with deformable convolutional layers, have been proposed to learn mappings between metasurface geometries and their corresponding spectral responses, thereby reducing computational costs and improving prediction accuracy~\cite{Waseer2025, dai2017deformable, zhu2019deformable}. However, the training of these models from scratch necessitates extensive datasets, which are often costly and time-intensive to generate. Moreover, these approaches may struggle to generalize well to unseen geometric configurations, particularly in scenarios with limited training data.} 

To address these limitations, transfer learning has emerged as a viable strategy, allowing pretrained networks to be fine-tuned for domain-specific tasks~\cite{howard2019searching}. MobileNetV2 \cite{sandler2018mobilenetv2}, a lightweight CNN architecture originally designed for image classification, has shown significant potential in feature extraction from grayscale images, making it a compelling candidate for spectral prediction in metasurface designs~\cite{sandler2018mobilenetv2}. By leveraging pretrained MobileNetV2 and adapting its architecture for spectral regression, the computational cost and training duration can be significantly reduced while retaining the network's feature extraction capabilities.

In addition to transfer learning, signal smoothing techniques, such as the Savitzky-Golay filter, offer further performance enhancements by mitigating noise in predicted spectra while preserving essential spectral features~\cite{savitzky1964smoothing}. The Savitzky-Golay filter applies polynomial fitting over a defined window length, effectively attenuating high-frequency noise and improving prediction stability. Additionally, the proposed model integrates a Multi-Channel Spectral Refinement (MCSR) module that employs multiple one-dimensional convolutional layers to further refine spectral predictions, capturing intricate spectral features that conventional CNNs may overlook.

\new{The main contributions of this work are summarized as follows. First, unlike prior methods that primarily rely on training convolutional architectures from scratch \cite{ding2023machine,Waseer2025}, the proposed framework integrates transfer learning, MCSR, and Savitzky--Golay smoothing to improve spectral stability and prediction accuracy under limited-data conditions for spectral prediction of MXene-based metasurface absorbers. Second, transfer learning with MobileNetV2 enables effective feature extraction under limited-data conditions. Third, the proposed MCSR module improves spectral prediction accuracy through localized spectral refinement in the one-dimensional spectral domain. Fourth, Savitzky--Golay smoothing enhances prediction stability while preserving physically meaningful spectral characteristics. Finally, comprehensive comparative and ablation analyses are conducted to evaluate the contribution of each component to the overall framework performance.}


\section{Methodology}
The proposed framework integrates a deep learning-based predictive model with Savitzky-Golay smoothing and MCSR to enhance the accuracy of spectral predictions for MXene-based metasurfaces. The methodology is structured into several key components, including dataset preparation, model architecture, spectral refinement, training protocol, and evaluation metrics.

\subsection{Dataset Preparation}

\begin{figure*}[ht!]
	\centering
	\includegraphics[width=0.8\linewidth]{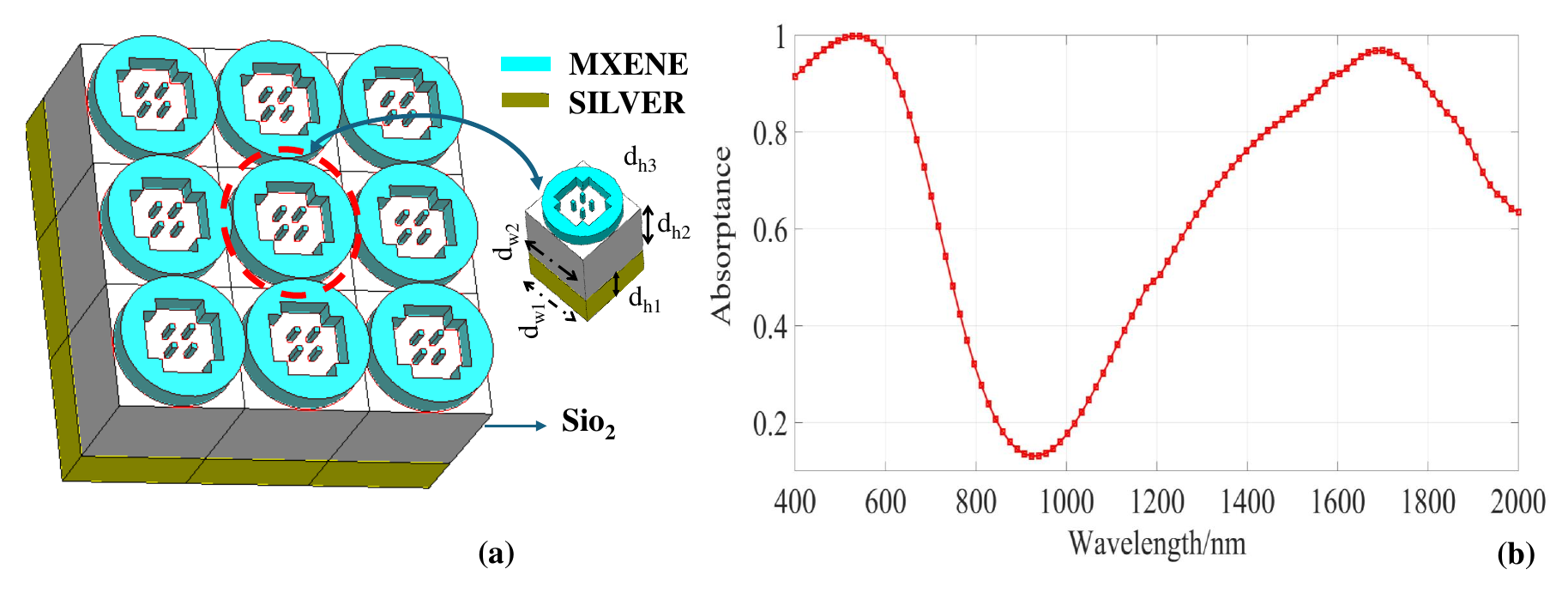}
	\caption{(a) Three-layer MIM metasurface-based absorber comprising a top MXene metasurface layer, middle SiO$_2$ dielectric substrate, and bottom Ag (silver) reflector. (b) Corresponding absorption spectra demonstrating dual-band absorption.}
	\label{fig1_description}
\end{figure*}

\new{The proposed absorber consists of a three-layer metal–insulator–metal (MIM) structure, comprising a patterned MXene metasurface as the top layer, a SiO$_2$ dielectric substrate as the intermediate layer, and a silver (Ag) reflective layer at the bottom, as illustrated in Fig.~\ref{fig1_description}. The MXene layer has a thickness of 40~nm and is responsible for engineered absorption through subwavelength patterning. The SiO$_2$ layer has a thickness of 200~nm, while the Ag back reflector has a thickness of 100~nm, ensuring negligible transmission. The lateral unit-cell dimension is $100 \times 100$~nm$^2$.}

\new{The material permittivity of MXene was taken from experimentally reported data~\cite{chaudhuri2018highly, mauchamp2014enhanced}, while SiO$_2$ relative permittivity was set to 2.25. The optical response of the metasurface was simulated using CST Microwave Studio, employing the finite-difference time-domain (FDTD) method. Periodic boundary conditions were applied in the $x$–$y$ plane, and open (radiation) boundary conditions were used along the $z$-direction. A normally incident plane wave was launched onto the top surface of the metasurface for excitation.}

\paragraph{\new{Dataset Generation.}}
\new{To construct the dataset, 500 distinct metasurface geometries were generated by varying the top-layer MXene patterns within a constrained parameter space. The variations include changes in feature dimensions (e.g., widths, radii, spacings, and structural layouts), while maintaining fixed layer thicknesses, unit-cell size, and illumination conditions. The design space was constrained to ensure physically realizable geometries compatible with fabrication considerations.}

\new{Additionally, symmetry-preserving transformations were applied to enhance geometric diversity, resulting in variations such as cross-, box-, and H-shaped patterns. This approach allows the dataset to capture a range of spatial configurations and corresponding spectral responses without altering the underlying physical constraints.}

\paragraph{\new{Spectral Simulation.}}
\new{For each metasurface design, the corresponding absorption spectrum was obtained through FDTD simulations in CST Studio Suite. The absorptivity \(A(\lambda)\) at wavelength \(\lambda\) was computed as:
\begin{equation}
A(\lambda) = 1 - R(\lambda) - T(\lambda),
\end{equation}
where $R(\lambda)$ and $T(\lambda)$ denote reflectance and transmittance, respectively. In terms of the scattering parameters, $R(\lambda)=|S_{11}|^2$ and $T(\lambda)=|S_{21}|^2$. Due to the sufficiently thick Ag reflector, transmission is negligible, i.e., \(T(\lambda) \approx 0\), leading to:
\begin{equation}
A(\lambda) \approx 1 - R(\lambda).
\end{equation}
The spectral response is sampled over 102 discrete wavelength points, resulting in a 102-dimensional absorption vector for each design.}

\paragraph{\new{Data Representation and Splitting.}}
\new{Each unit-cell geometry was converted into a grayscale image representation and resized to $64 \times 64$ pixels, with the corresponding absorption spectrum represented as a 102-element vector, for deep learning-based setup. The dataset was then partitioned into training, validation, and test sets using an 80:10:10 split. To ensure statistical robustness, multiple independent random splits were used during evaluation, and performance metrics are averaged across runs. This dataset construction strategy ensures sufficient geometric diversity while maintaining a controlled design space, enabling effective training and evaluation of the proposed model under realistic and computationally feasible conditions.}

\subsection{Model Architecture}
\begin{figure*}[ht]
	\centering
	\includegraphics[width=1\linewidth]{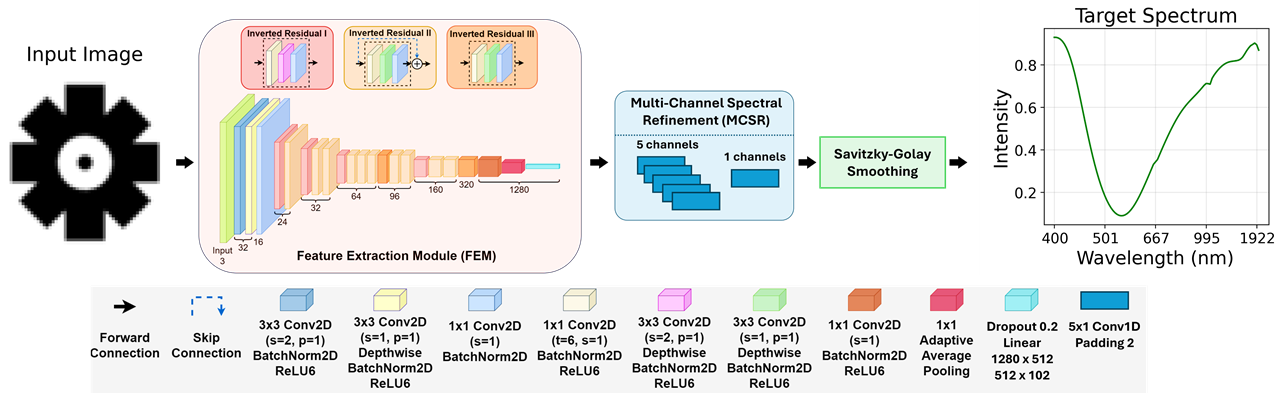}
	\caption{\new{Proposed framework integrating a pretrained MobileNetV2 backbone, Multi-Channel Spectral Refinement, and Savitzky--Golay post-processing for spectral prediction.}}
    
	\label{fig:arch}
\end{figure*}

The proposed model depicted in Figure~\ref{fig:arch} leverages MobileNetV2~\cite{sandler2018mobilenetv2}, a lightweight deep convolutional neural network that is originally pretrained on ImageNet\new{, as a feature extraction backbone}. 

Although MobileNetV2 is pretrained on ImageNet using RGB images, the learned low-level features such as edges, textures, and contours are largely domain-agnostic and transferable to grayscale inputs. \newr{In our implementation, grayscale metasurface images are adapted to the expected input format using channel replication and the network is fine-tuned end-to-end, allowing higher-level features to specialize for the spectral prediction task.} 

The MobileNetV2 backbone is modified for spectral regression by replacing the \new{original classification head with fully connected projection layers followed by} an MCSR block. 

\new{A fixed parameter-based Savitzky-Golay smoothing is applied as a post-processing step to further refine the predicted spectra.}

\new{For \emph{Feature Extraction Module (FEM)}, the MobileNetV2 backbone is adapted to accept grayscale input images and is used to extract spatial features from the metasurface geometry. The backbone outputs a feature vector of dimension \(1280\), which is subsequently projected into a lower-dimensional latent representation using a fully connected layer:
\(
\mathbf{h} \in \mathbb{R}^{512}.
\)
A second fully connected layer then maps this latent representation to a \(102\)-dimensional vector, corresponding to the predicted absorption spectrum:
\(
\mathbf{y}_0 \in \mathbb{R}^{102}
\).}

\new{The initial predicted spectrum vector is further refined using the \emph{MCSR} module, which consists of two one-dimensional convolutional layers with kernel size 5 and padding 2. This design introduces an explicit spectral-domain refinement stage that is decoupled from spatial feature extraction, enabling targeted correction of prediction errors at different spectral scales, which is not explicitly modeled in standard end-to-end CNN regression pipelines. The first layer expands the single-channel spectral representation into multiple channels (1 $\rightarrow$ 5), while the second layer projects it back to a single-channel spectrum (5 $\rightarrow$ 1). No pooling is applied, preserving the spectral resolution.} \new{This design enables localized spectral refinement of spectral features, improving the model’s ability to capture fine spectral variations.}

\new{Unlike prior approaches that rely solely on direct regression from spatial features, the proposed framework introduces an intermediate spectral refinement stage, allowing the model to explicitly model and correct spectral inconsistencies.}

To further improve the performance, the output of the MCSR
\(\mathbf{y}_{\mathrm{ref}}\)
is subsequently processed using \emph{Savitzky-Golay Smoothing}\new{, applied as a post-processing step to the predicted spectrum. In contrast to the MCSR module, this operation is applied only at inference. The filter performs local polynomial fitting over a sliding window and smooths the spectrum through one-dimensional convolution with fixed coefficients. It is applied to the MCSR predicted output, along the spectral dimension} to reduce spectral noise.

\new{The implementation of the proposed framework is publicly available at: [due to double blind review, link will be shared after acceptance \url{https://github.com/eeshahid/MXene_spectral_prediction}.}

\subsection{\new{Savitzky-Golay Smoothing}}

The Savitzky-Golay filter is a digital smoothing filter widely employed in signal processing to reduce noise while preserving the \new{low-order structure} of the signal, such as peaks, troughs, and overall trends \cite{savitzky1964smoothing}. \new{Unlike simple averaging filters, it performs local polynomial fitting within a sliding window, enabling effective suppression of high-frequency variations without strongly distorting local low-order polynomial trends. The filter operates by fitting a polynomial to a local window of data points using the method of least squares and evaluating the fitted polynomial at the central point.}

\new{While Savitzky-Golay filtering is a classical signal processing technique, its integration as a post-processing step within a deep spectral prediction pipeline is designed to complement learned representations by explicitly enforcing smoothness and reducing high-frequency prediction artifacts.}

\subsubsection{\new{Window Length and Polynomial Order}}

\new{Two parameters govern the behavior of the filter:
\begin{itemize}
    \item Window length (\(W\)): The number of samples used for local polynomial fitting. The window length should be an odd integer to ensure symmetry about the central point. In this work, \(W = 11\), corresponding to a half-window size of
    \(n = \frac{W - 1}{2}\).
    \item Polynomial order (\(d\)): The degree of the polynomial used to approximate the signal within each window. In this work, \(d = 2\), corresponding to quadratic fitting.
\end{itemize}}

\subsubsection{\new{Filter Formulation}}

\new{Let \(\mathbf{y}_{\mathrm{ref}}\) denote the refined predicted spectrum, and let \(y_i\) be its \(i\)-th spectral sample. For each spectral position \(i\), a local window of values
\(\{y_{i-n}, \ldots, y_{i}, \ldots, y_{i+n}\}\)
is considered. A polynomial of degree \(d\) is fitted to these samples by minimizing the least-squares error. The smoothed value at position \(i\) is then obtained by evaluating the fitted polynomial at the central point.}

\new{Formally, let \(x_k = k\), for \(k = -n, \ldots, n\), denote the local coordinate system centered at the current sample. A design matrix \(A \in \mathbb{R}^{W \times (d+1)}\) is constructed as:}
\[
\new{
A =
\begin{bmatrix}
1 & x_{-n} & x_{-n}^2 & \cdots & x_{-n}^d \\
\vdots & \vdots & \vdots & \ddots & \vdots \\
1 & x_{0} & x_{0}^2 & \cdots & x_{0}^d \\
\vdots & \vdots & \vdots & \ddots & \vdots \\
1 & x_{n} & x_{n}^2 & \cdots & x_{n}^d
\end{bmatrix}
}
\]

\new{Assuming that \(A\) has full column rank, the least-squares solution is obtained using the normal equations, yielding the pseudo-inverse:}
\begin{equation}
\new{
A^{+} = (A^T A)^{-1} A^T
}
\end{equation}

\new{The Savitzky-Golay filter coefficients correspond to evaluating the fitted polynomial at the central point \(x=0\). This is achieved by selecting the row associated with the constant term, leading to the filter kernel:}
\begin{equation}
\new{
\mathbf{c} = e_0^T (A^T A)^{-1} A^T
}
\end{equation}
\new{where \(e_0 = [1, 0, \ldots, 0]^T\). The entries of \(\mathbf{c}\) define the filtering coefficients, i.e., \(\mathbf{c} = [c_{-n}, \ldots, c_n]\)}.

\new{This procedure is equivalent to applying a linear filtering operation with a set of fixed coefficients:}
\begin{equation}
\new{
\hat{y}_i = \sum_{k=-n}^{n} c_k \, y_{\mathrm{ref},\, i+k},
}
\end{equation}
\new{where \(c_k\) are the Savitzky--Golay filter coefficients determined by the chosen \(W\) and \(d\).
Applying this operation across all spectral indices yields the smoothed spectrum:}
\begin{equation}
\new{
\hat{\mathbf{y}} = \mathbf{c} * \mathbf{y}_{\mathrm{ref}},
}
\end{equation}
\new{where \(*\) denotes one-dimensional filtering/convolution with appropriate boundary handling, and \(\mathbf{c} = [c_{-n}, \ldots, c_n]\) represents the fixed filter kernel.
}
\subsubsection{\new{Smoothing in the Proposed Framework}}

\new{In the proposed model, Savitzky-Golay smoothing is applied to the refined predicted spectrum \(\mathbf{y}_{\mathrm{ref}}\) produced by the MCSR module. The filter operates along the spectral dimension (102-point vector) and produces the final output spectrum:}
\begin{equation}
\new{
\hat{\mathbf{y}} = \mathrm{SGFilter}(\mathbf{y}_{\mathrm{ref}}; W=11, d=2).
}
\end{equation}
\new{The filter preserves low-order polynomial trends within each window while attenuating higher-frequency variations. At the spectral boundaries, appropriate padding or boundary extension is used to maintain the output length. The selected parameters provide a balance between noise suppression and preservation of the underlying spectral structure, as further validated through the sensitivity analysis presented in Section~\ref{sec:abl_savgol}.}

\subsection{Training Protocol}
\begin{figure*}[h]
	\centering
	\includegraphics[width=0.9\linewidth]{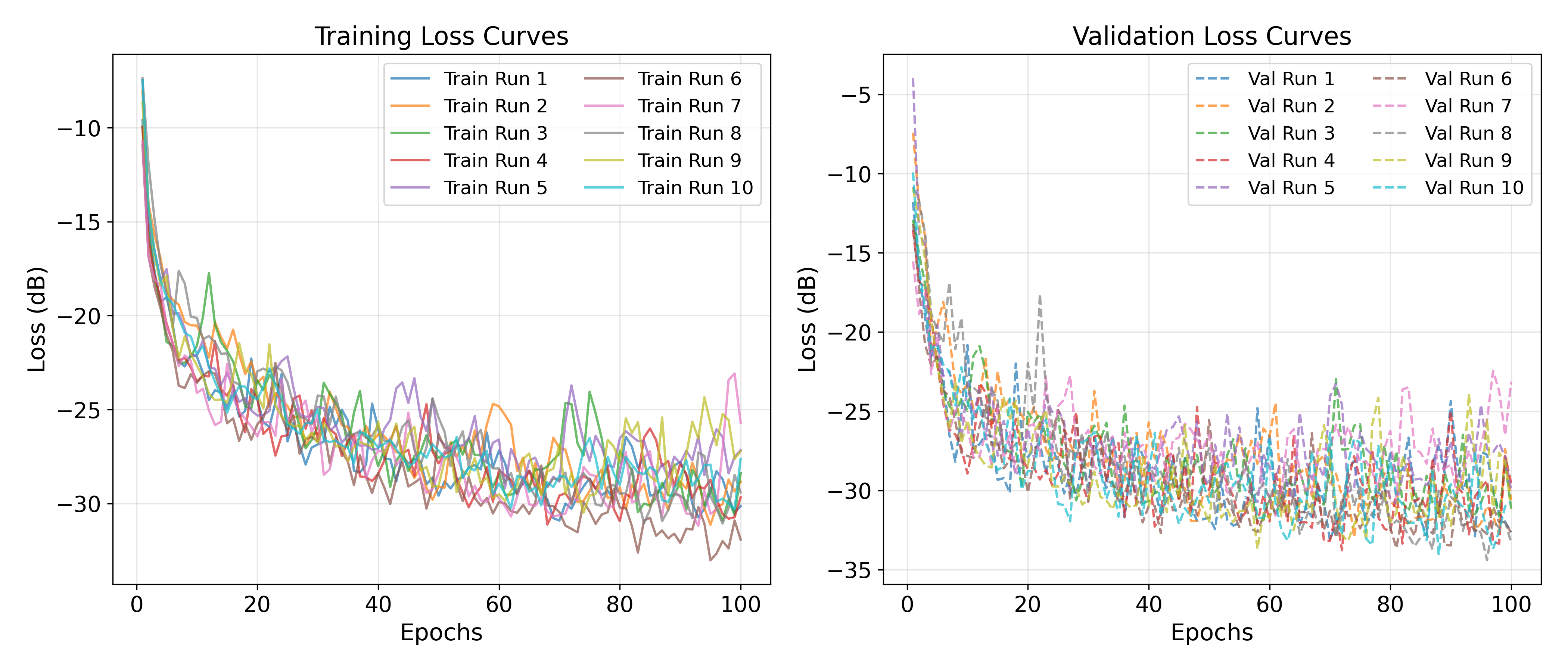}
    \includegraphics[width=0.9\linewidth]{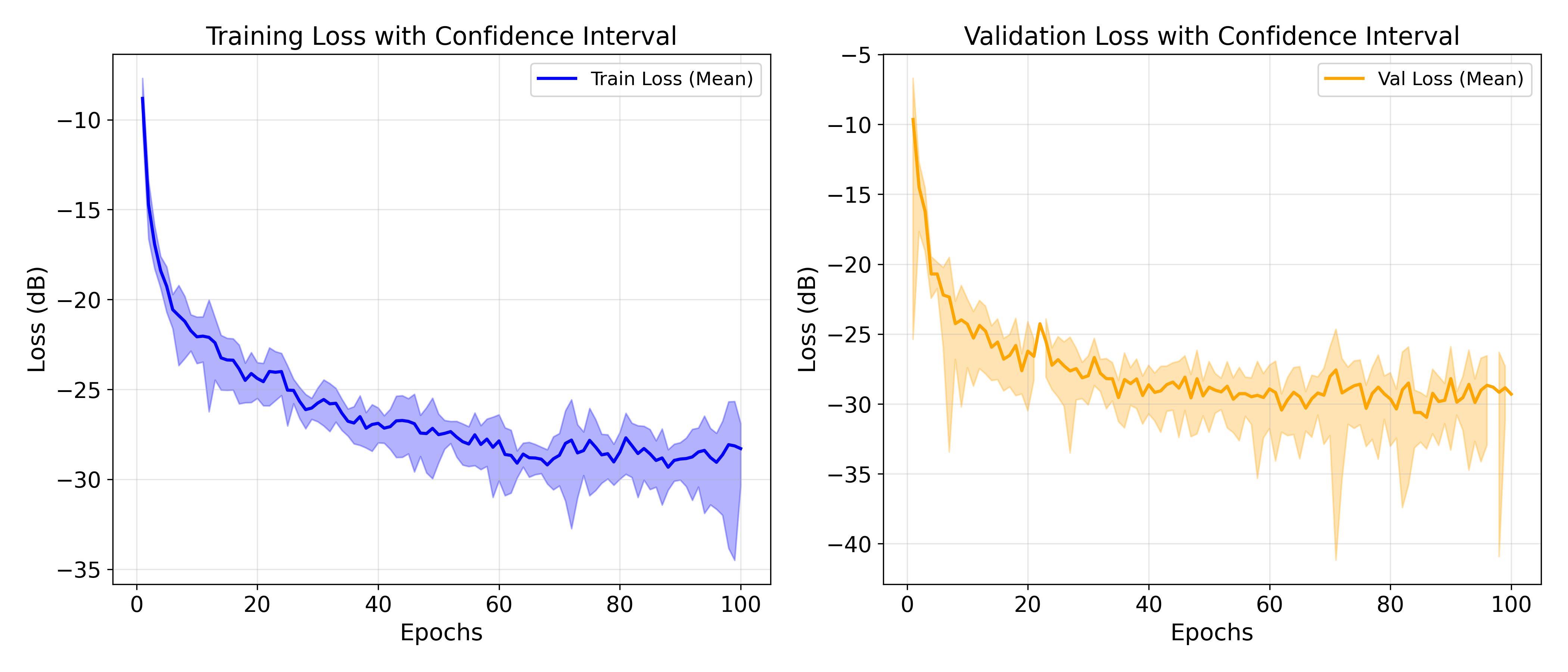}
    \caption{\new{Training (top left) and validation (top right) loss curves plotted as \(10 \log_{10}(\text{MSE})\) (in dB) for the proposed model. The bottom plots illustrate confidence intervals across multiple runs, demonstrating training stability and consistent performance.}}
    \label{figloss}
\end{figure*}

\new{The model is trained using the Adam optimizer~\cite{kingma2014adam} with a learning rate of \(1\times10^{-3}\) and a batch size of 32. The loss function is the Mean Squared Error (MSE), defined as:}
\begin{equation}
\mathcal{L} = \frac{1}{N} \sum_{i=1}^{N} \left( y_{\mathrm{ref},i} - y_i \right)^2,
\end{equation}
where \( y_{\mathrm{ref},i} \) and \( y_i \) denote the predicted spectral value after MCSR block (before Savitzky--Golay post-processing) and ground truth spectral values, respectively.


\new{The network is trained end-to-end using the backbone feature extractor, fully connected regression head, and MCSR module. Savitzky--Golay smoothing is not applied during training and is instead used as a post-processing step during evaluation at inference time.}

\new{To prevent overfitting, dropout with a rate of 0.2 is applied in the fully connected layers. Early stopping is employed based on validation loss with a patience of 50 epochs. All convolutional layers are initialized using He initialization~\cite{he2015delving, glorot2010understanding} to ensure stable gradient propagation.}

\new{For statistical robustness, all experiments are repeated over 10 independent runs. In each run, the dataset is randomly split into training (80\%), validation (10\%), and testing (10\%) subsets using different random seeds. The reported results correspond to the mean and standard deviation across these runs. The training and validation loss curves, along with confidence intervals across runs, are shown in Figure~\ref{figloss}, demonstrating stable convergence and low variance across different data splits. The same training protocol is applied consistently across all baseline models to ensure fair comparison.}


\subsection{Evaluation Metrics}

\new{Model performance is evaluated using three complementary metrics: root mean squared error (RMSE), coefficient of determination (\(R^2\)), and peak signal-to-noise ratio (PSNR). These metrics collectively assess prediction accuracy, variance explanation, and signal reconstruction fidelity for the predicted spectra.}

\new{The RMSE is defined as}
\begin{equation}
\new{
\text{RMSE} = \sqrt{\frac{1}{N} \sum_{i=1}^{N} (\hat{y}_i - y_i)^2},
}
\end{equation}

\new{where \(\hat{y}_i\) denote the final predicted spectrum after Savitzky--Golay post-processing, and \(y_i\) denote the ground truth spectral values. RMSE measures the average magnitude of prediction error and provides an interpretable estimate of absolute deviation in the spectral domain.}

\new{The coefficient of determination (\(R^2\)) is given by}
\begin{equation}
\new{
R^2 = 1 - \frac{\sum_{i=1}^{N} (\hat{y}_i - y_i)^2}{\sum_{i=1}^{N} (y_i - \bar{y})^2},
}
\end{equation}
\new{where \(\bar{y}\) represents the mean of the ground truth spectrum. This metric quantifies the proportion of variance explained by the model and provides a normalized measure of goodness-of-fit across different spectral samples.}

\new{The PSNR is defined as}
\begin{equation}
\new{
\text{PSNR} = 10 \log_{10} \left( \frac{y_{\max}^2}{\text{MSE}} \right),
}
\end{equation}
\new{where \(\text{MSE} = \frac{1}{N} \sum_{i=1}^{N} (\hat{y}_i - y_i)^2\) and \(y_{\max}\) denotes the maximum possible value of the spectrum. In this work, the absorption spectrum is a bounded continuous signal within the range \([0,1]\), making PSNR a meaningful indicator of signal fidelity. Unlike RMSE, which captures absolute error, PSNR emphasizes the relative magnitude of reconstruction error with respect to the signal dynamic range. This is particularly important for evaluating the preservation of fine spectral structures, such as narrowband peaks and local variations, which are critical for accurate metasurface characterization.}

\new{Together, RMSE, \(R^2\), and PSNR provide a comprehensive evaluation framework that captures absolute error, relative explanatory power, and signal-level fidelity, respectively, enabling robust assessment of spectral prediction performance.}

\subsection{Explainability via Grad-CAM}

\new{To interpret the model's predictive focus, Gradient-weighted Class Activation Mapping (Grad-CAM)~\cite{selvaraju2017grad} is applied to the final convolutional layer of the backbone network}, generating heatmaps that visualize the spatial regions in the input metasurface geometry most influential to the predicted spectral response.
These visualizations provide insights into how specific geometrical features affect spectral characteristics\new{, enhancing the interpretability of the proposed model}.



\section{Results and Discussion}

\subsection{Training Performance Analysis}

The training performance of the proposed framework was evaluated by monitoring the loss convergence across 100 epochs. The learning curves for both training and validation sets are depicted in Fig.~\ref{figloss}.
\new{The model exhibits rapid convergence during the initial training phase, followed by gradual stabilization, indicating effective optimization. The close alignment between training and validation loss curves suggests minimal overfitting and good generalization capability. Furthermore, the narrow confidence intervals across multiple runs demonstrate consistent convergence behavior and low sensitivity to data splits. These observations indicate that the proposed model with combination of transfer learning and MCSR enables stable training even in a relatively data-limited setting.}

\begin{figure}[!htbp]
	\centering
        \includegraphics[width=0.9\textwidth]{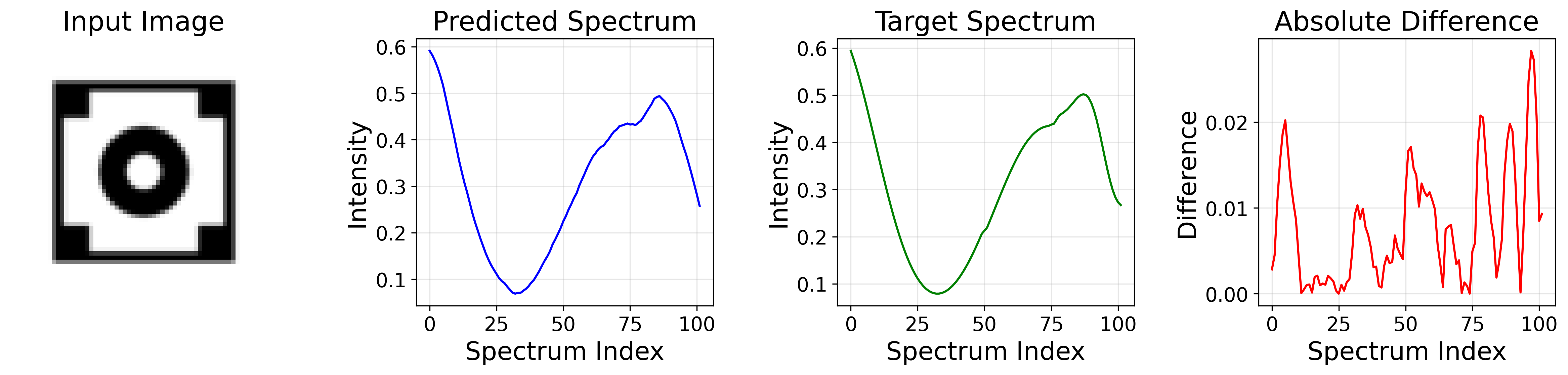}
        \includegraphics[width=0.9\textwidth]{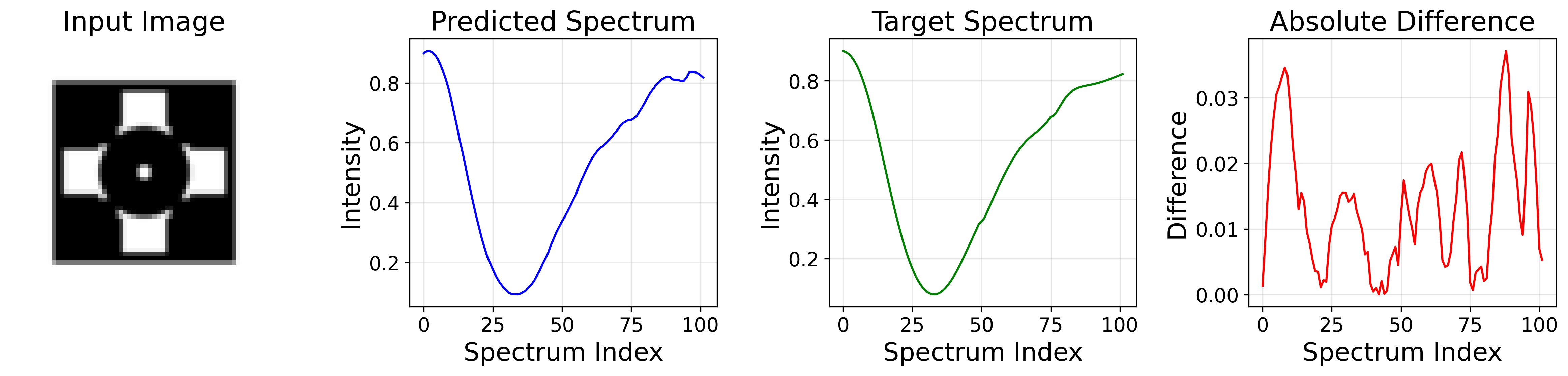}
        \includegraphics[width=0.9\textwidth]{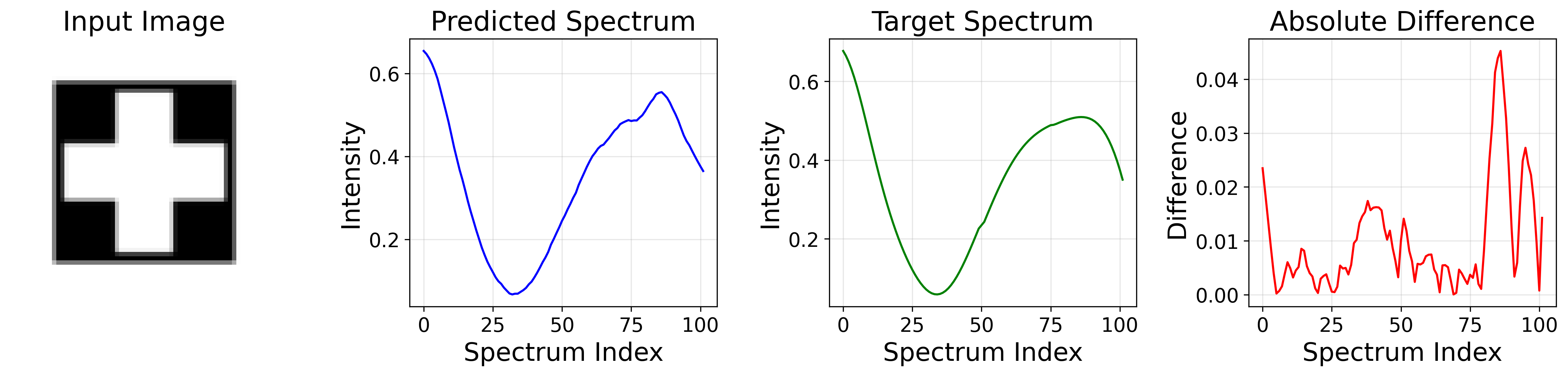}
        \includegraphics[width=0.9\textwidth]{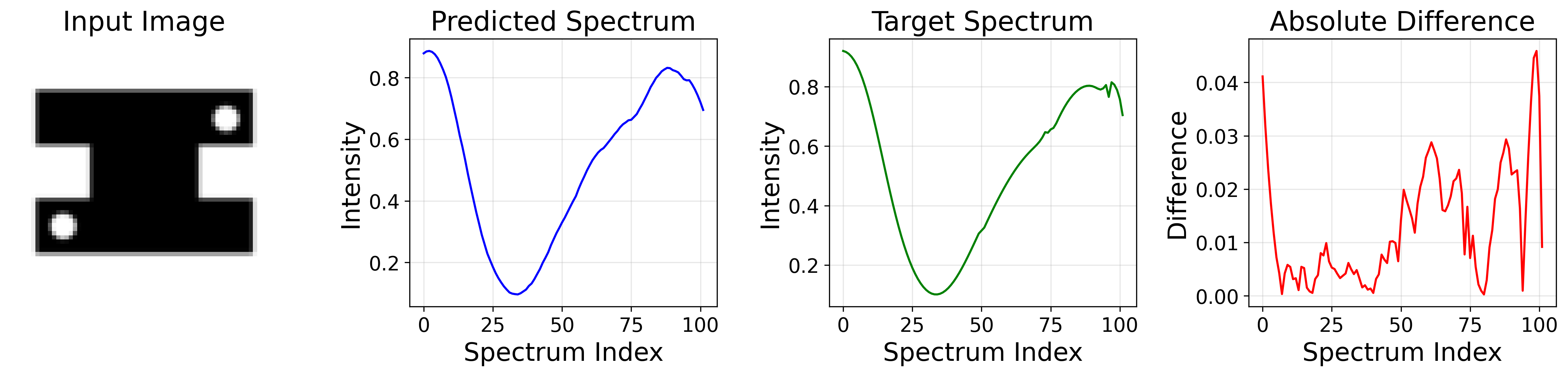}
        \includegraphics[width=0.9\textwidth]{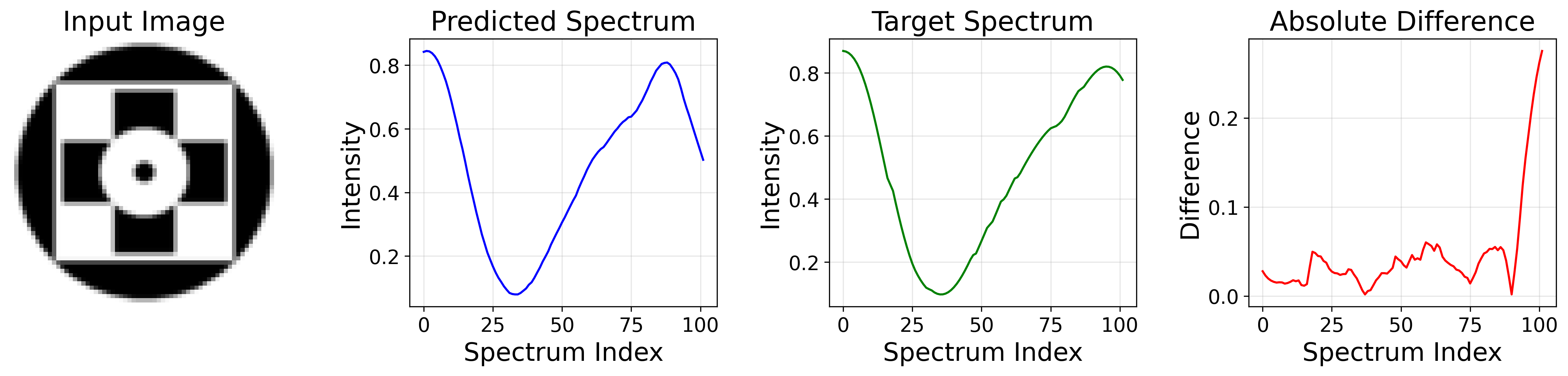}
        \includegraphics[width=0.9\textwidth]{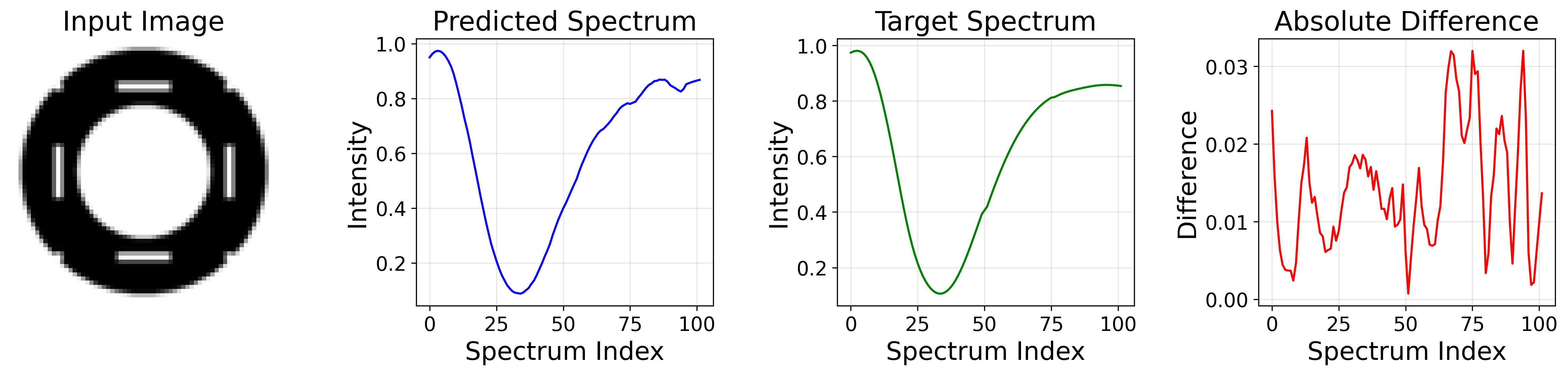}
	\caption{Comparison of the predicted vs. actual spectral \new{\textbf{absorptance}} for six representative metasurface designs. The model accurately captures both broadband and narrowband absorption features, benefiting from the MCSR layer and smoothing post-processing.}
	\label{fig4}
\end{figure}

\subsection{Qualitative Evaluation of Spectral Predictions}
Qualitative evaluation was performed by comparing the predicted spectra against the ground truth spectra obtained through CST simulations. Six representative metasurface designs were selected, demonstrating varying spectral responses to assess the model's predictive capability (Fig.~\ref{fig4}). 

\new{The proposed model effectively captures the overall spectral shape, including the absorption peaks and troughs across diverse metasurface geometries.} The use of transfer learning and MCSR notably enhanced the spectral resolution, reducing deviations from the ground truth.
\new{The corresponding absolute error plots in Fig.~\ref{fig4} further indicate that most prediction deviations remain relatively small and spatially localized along the spectral axis. For broadband solar harvesting applications, these localized deviations are unlikely to significantly alter the overall absorbed energy distribution. However, for resonance-sensitive applications such as refractive-index sensing, localized prediction errors near sharp spectral dips may influence sensitivity estimation and figure-of-merit calculations, highlighting the importance of localized spectral accuracy.}

\subsection{Quantitative Evaluation and Model Performance}



\new{The quantitative evaluation was conducted using RMSE, \(R^2\), and PSNR across 10 independent runs, as summarized in Table~\ref{tab:proposed_all_runs_original_order}. The proposed model achieves an average RMSE of 0.0227, \(R^2 = 0.9563\), and PSNR of 33.10 dB. The relatively low standard deviation across runs indicates consistent performance and robustness to variations in data splits.}

\new{Notably, the variations in RMSE and \(R^2\) across runs remain limited, suggesting that the model does not rely on specific training subsets and instead learns stable and generalizable representations. This robustness is particularly important in the context of limited training data, where model variance can otherwise be significant.}

\begin{table}[]
\centering
\caption{\new{Quantitative per-run performance of the proposed model over 10 independent runs. Results are reported on the test set, with the final row showing the mean and standard deviation across runs.}}
\resizebox{0.8\linewidth}{!}{
\begin{tabular}{c|c|c|c}
\hline \hline
\textbf{{\diagbox[width=14em]{Run \#}{Metric}}} & \textbf{RMSE} $\downarrow$ & \textbf{R$^2$} $\uparrow$ & \textbf{PSNR (dB)} $\uparrow$ \\
\hline

1 & 0.0184 & 0.9652 & 34.7111 \\
2 & 0.0273 & 0.9665 & 31.2908 \\
3 & 0.0242 & 0.9743 & 32.3357 \\
4 & 0.0164 & 0.9656 & 35.6922 \\
5 & 0.0178 & 0.9804 & 35.0051 \\
6 & 0.0216 & 0.9503 & 33.2925 \\
7 & 0.0178 & 0.9624 & 34.9792 \\
8 & 0.0333 & 0.9245 & 29.5494 \\
9 & 0.0255 & 0.9386 & 31.8635 \\
10 & 0.0243 & 0.9353 & 32.2704 \\

\hline
Mean $\pm$ Std &
0.0227 $\pm$ 0.0050 & 0.9563 $\pm$ 0.0174 & 33.0990 $\pm$ 1.8760
\\ \hline \hline
\end{tabular}
}
\label{tab:proposed_all_runs_original_order}
\end{table}

\subsection{Comparison with State-of-the-Art Methods}

\new{To assess the efficacy of the proposed approach, a comparative analysis was conducted against baseline conventional CNN and deformable CNN models \cite{Waseer2025}, as shown in Table~\ref{tab:comparison}. The baseline models are implemented as reported, and evaluated under the same experimental setup as the proposed model. Specifically, all models are trained using identical dataset, training/validation/test splits (80/10/10), and protocol, with results reported as mean $\pm$ standard deviation over 10 independent runs. The proposed model achieves a 34.6\% reduction in RMSE, a 6.1\% increase in \(R^2\), and a 3.53 dB improvement in PSNR compared to the deformable CNN baseline.
In addition to these gains in average performance, the proposed model consistently exhibits lower standard deviation across all metrics, indicating improved stability and robustness to variations in data splits across runs, suggesting that the proposed framework not only improves prediction accuracy but also enhances generalization and robustness in the data-limited regime.}

\new{Recent transformer-based approaches for spectral prediction offer strong modeling capacity but often require large-scale datasets and higher computational resources. In contrast, the proposed framework leverages transfer learning and a lightweight convolutional design to achieve competitive performance under limited data conditions. Furthermore, while inverse-design methods typically focus on iterative or generative optimization of geometries, the proposed approach targets efficient forward spectral prediction with explicit refinement mechanisms.}


\begin{table}[]
\centering
\caption{\new{Comparison of the proposed model with baseline CNN and deformable CNN models. All models are evaluated on the same dataset and data splits, with results reported as mean $\pm$ standard deviation over 10 runs.}}
\resizebox{0.8\linewidth}{!}{
\begin{tabular}{l|c|c|c}
\hline \hline
\textbf{{\diagbox[width=14em]{Model}{Metric}}} & \textbf{RMSE} $\downarrow$ & \textbf{R$^{2}$ $\uparrow$}  & \textbf{PSNR (dB)} $\uparrow$ \\
\hline
CNN (Baseline)~\cite{Waseer2025}  & 0.0430 $\pm$ 0.0133 & 0.8883 $\pm$ 0.0712 & 27.86 $\pm$ 2.80 \\
Deformable CNN (Baseline)~\cite{Waseer2025} & 0.0347 $\pm$ 0.0136 & 0.9010 $\pm$ 0.0685 & 29.57 $\pm$ 3.40 \\
Proposed &
0.0227 $\pm$ 0.0050 & 0.9563 $\pm$ 0.0174 & 33.10 $\pm$ 1.88
\\ \hline \hline
\end{tabular}
}
\label{tab:comparison}
\end{table}

\new{These results suggest that the proposed framework benefits not only from improved feature extraction but also from the interaction between learned spectral representations and structured post-processing, highlighting the importance of system-level design in spectral prediction tasks.}

\new{The improved performance of the proposed framework can be attributed to the complementary interaction between transfer learning, spectral refinement, and smoothing. The pretrained MobileNetV2 backbone provides robust feature representations under limited data conditions, while the proposed MCSR module explicitly refines localized spectral inconsistencies that are not effectively captured by conventional CNN-based regression architectures. In addition, the Savitzky–Golay smoothing improves spectral stability by suppressing high-frequency fluctuations while preserving important resonance characteristics. The ablation study further confirms that each component contributes incrementally to the overall performance improvement.}

\subsection{\new{Ablation Study and Sensitivity Analysis}}

\new{This ablation study is designed to evaluate the contribution of each component and to validate that the performance gains arise from their combined effect rather than any single module in isolation.}

\new{To systematically evaluate the contribution of each component in the proposed framework, we conduct a comprehensive ablation study covering: (i) architectural components (transfer learning, MCSR, smoothing), (ii) backbone selection, (iii) training data size, and (iv) Savitzky-Golay hyperparameters.}

\subsubsection{\new{Effect of Model Architecture Components}}

\newr{This ablation study systematically evaluates different combinations of transfer learning, Multi-Channel Spectral Refinement (MCSR), and Savitzky--Golay smoothing to quantify the individual and combined contribution of each component. The results, summarized in Table~\ref{tab:ablation_study}, demonstrate how each module affects model performance in terms of RMSE, \(R^2\), and PSNR.}

\new{The baseline configuration, consisting solely of smoothing without transfer learning or MCSR, yields the highest RMSE of \( 0.0430 \pm 0.0133 \), the lowest \( R^2 \) score of \( 0.8883 \pm 0.0712 \), and a PSNR of \( 27.86 \pm 2.80 \), indicating that smoothing alone is insufficient to recover accurate spectral predictions and primarily acts as a post-processing refinement rather than a core predictive component.}

\newr{We further evaluate the combination of MCSR and smoothing without transfer learning. This configuration achieves only marginal improvement over smoothing alone (RMSE \(0.0399 \pm 0.0065\) versus \(0.0430 \pm 0.0133\)), indicating that MCSR alone is insufficient to compensate for the absence of pretrained feature representations. This observation highlights that effective spatial feature extraction through transfer learning is a prerequisite for the proposed spectral refinement strategy.}

\new{Introducing transfer learning alone (without MCSR and smoothing) improves performance, reducing RMSE to \(0.0347 \pm 0.0136\) and increasing PSNR to \(29.57 \pm 3.40\), demonstrating the effectiveness of pretrained feature extraction in improving generalization under limited data conditions.}

\new{Adding MCSR on top of transfer learning (without smoothing) results in further improvements, with RMSE decreasing to \(0.0322 \pm 0.0126\) and PSNR increasing to \(30.29 \pm 3.08\), highlighting the role of MCSR in refining spectral representations and capturing localized spectral patterns.}

\new{Next, we evaluate the combination of transfer learning and smoothing, without MCSR. In this case, the RMSE reduces to \(0.0310 \pm 0.0101\), and PSNR improves to \(30.58 \pm 2.76\), indicating that smoothing effectively reduces high-frequency prediction noise even without explicit spectral refinement.}

\newr{The fully integrated configuration, combining transfer learning, MCSR, and Savitzky--Golay smoothing, achieves the overall best performance.}

\new{Importantly, the improvement obtained with Savitzky-Golay smoothing is not merely due to trivial smoothing effects, as the ablation results show consistent gains across RMSE, \(R^2\), and PSNR without degrading spectral structure, indicating improved stability rather than loss of detail.}

\new{Overall, the results confirm that optimal performance is achieved through the synergistic integration of all three components, with each module addressing a distinct aspect of the spectral prediction task.}

\begin{table*}[]
\centering
\caption{\new{Ablation study on the effect of multi-channel spectral refinement, transfer learning, and Savitzky-Golay smoothing components on model performance}}
\resizebox{0.9\linewidth}{!}{
\begin{tabular}{c c c|c c c}
\hline \hline
\textbf{Transfer Learning} & \textbf{MCSR} & \textbf{Smoothing} & \textbf{RMSE} ↓ & \textbf{R$^{2}$ Score} ↑ & \textbf{PSNR (dB)} ↑ \\
\hline



\ding{55} & \ding{55} & \ding{51} & 0.0430 $ \pm $ 0.0133 & 0.8883 $ \pm $ 0.0712 & 27.86 $ \pm $ 2.80 \\


\newr{\ding{55}} & \newr{\ding{51}} & \newr{\ding{51}} & \newr{0.0399 $\pm$ 0.0065} & \newr{0.9003 $\pm$ 0.0283} & \newr{28.10 $\pm$ 1.38} \\ 

\ding{51} & \ding{55} & \ding{55} & 0.0347 $ \pm $ 0.0136 & 0.9010 $ \pm $ 0.0685 & 29.57 $ \pm $ 3.40 \\
\ding{51} & \ding{51} & \ding{55} & 0.0322 $ \pm $ 0.0126 & 0.9031 $ \pm $ 0.0754 & 30.29 $ \pm $ 3.08 \\
\ding{51} & \ding{55} & \ding{51} & 0.0310 $ \pm $ 0.0101 & 0.9051 $ \pm $ 0.0548 & 30.58 $ \pm $ 2.76 \\

\ding{51} & \ding{51} & \ding{51} &
0.0227 $\pm$ 0.0050 & 0.9563 $\pm$ 0.0174 & 33.10 $\pm$ 1.88
\\ \hline \hline
\end{tabular}
}
\label{tab:ablation_study}
\end{table*}

\subsubsection{\new{Effect of Feature Extraction Backbones}}

\new{To further validate the effectiveness of the MobileNetV2 backbone in the proposed model, we compare it against several alternative architectures, including Visual Geometry Group( VGG) \cite{simonyan2014very}, EfficientNet \cite{tan2019efficientnet}, Convolutional Next (ConvNeXt) \cite{liu2022convnet}, and Vision Transformer (ViT)-based \cite{dosovitskiy2020image} models, all initialized with ImageNet-pretrained weights. In addition, we include two lightweight baselines: (i) a simple CNN composed of three convolutional blocks (Conv–BatchNorm–ReLU with average pooling) followed by global pooling and a linear projection, and (ii) a fully connected Multi-Layer Perceptron (MLP) that operates on flattened input images with two hidden layers (1024 and 512 units) and dropout regularization.}

\new{The results in Table~\ref{tab:abl_backbones} show that the proposed MobileNetV2-based model achieves the best overall performance, with RMSE of \(0.0227 \pm 0.0050\) and \(R^2 = 0.9563 \pm 0.0174\), outperforming both lightweight baselines and significantly larger architectures such as VGG-11-BN and VGG-16-BN. Notably, while deeper and transformer-based models (e.g., ViT and ConvNeXt) offer higher representational capacity, they do not yield improved performance in this data-limited setting and, in some cases, exhibit instability, as reflected by higher variance across runs.}

\new{In contrast, MobileNetV2 provides an effective balance between feature extraction capability, parameter efficiency, and generalization performance, making it particularly well-suited for the proposed spectral prediction task.}

\begin{table*}[]
\centering
\caption{\new{Comparison of Proposed Model with different feature extraction backbone architectures}}
\resizebox{0.85\linewidth}{!}{
\begin{tabular}{l|c|c|c}
\hline \hline
\textbf{{\diagbox[width=14em]{Backbone}{Metric}}} & \textbf{RMSE} $\downarrow$ & \textbf{R$^{2}$ Score $\uparrow$}  & \textbf{PSNR (dB)} $\uparrow$ \\
\hline
Simple-CNN & 0.0596 $\pm$ 0.0118 & 0.8873 $\pm$ 0.0371 & 24.6464 $\pm$ 1.5751 \\
Simple-MLP & 0.0323 $\pm$ 0.0066 & 0.9221 $\pm$ 0.0372 & 29.9993 $\pm$ 1.8445 \\
VGG-11-BN \cite{simonyan2014very} & 0.0349 $\pm$ 0.0157 & 0.9142 $\pm$ 0.1259 & 29.8337 $\pm$ 3.2712 \\
VGG-16-BN \cite{simonyan2014very} & 0.0347 $\pm$ 0.0111 & 0.9485 $\pm$ 0.0166 & 29.6015 $\pm$ 2.5940 \\
EfficientNet-B0 \cite{tan2019efficientnet} & 0.0296 $\pm$ 0.0046 & 0.9214 $\pm$ 0.0544 & 30.6757 $\pm$ 1.3098 \\
ConvNeXt-Tiny \cite{liu2022convnet} & 0.0463 $\pm$ 0.0555 & 0.7600 $\pm$ 0.4395 & 31.1151 $\pm$ 7.7938 \\
ViT-B/16 \cite{dosovitskiy2020image} & 0.0322 $\pm$ 0.0101 & 0.9361 $\pm$ 0.0599 & 30.2781 $\pm$ 2.7122 \\
MobileNet-V2 (proposed) & 0.0227 $\pm$ 0.0050 & 0.9563 $\pm$ 0.0174 & 33.0990 $\pm$ 1.8760
\\ \hline \hline
\end{tabular}
}
\label{tab:abl_backbones}
\end{table*}

\subsubsection{\new{Effect of Training Data Size on Model Performance and Robustness}}

\new{To analyze the robustness of the proposed model under limited data conditions, we evaluate performance using different fractions of the training set, as summarized in Table~\ref{tab:abl_train_frac}.
As expected, performance generally improves with increasing training data. The RMSE decreases from \(0.0361 \pm 0.0107\) at 50\% training data to \(0.0227 \pm 0.0050\) when the full training set is used, with a corresponding increase in \(R^2\) score from \(0.8892\) to \(0.9563\). This trend indicates that the model benefits from additional training samples, improving both accuracy and stability. Some fluctuations are observed at intermediate fractions (e.g., 0.8), which can be attributed to statistical variability arising from limited sample sizes and random partitioning. However, the overall trend remains consistent across runs, as reflected by the results.}

\new{Importantly, even with reduced training data (e.g., 50\%–70\%), the model maintains competitive performance, demonstrating the effectiveness of transfer learning in mitigating data scarcity.
Furthermore, the relatively low degradation at reduced data fractions further indicates that the model does not rely on memorization but learns generalizable feature representations.
This suggests that the proposed framework can generalize reasonably well even in data-limited scenarios, which is critical for practical nanophotonic design tasks where dataset generation is computationally expensive.}

\begin{table*}[]
\centering
\caption{\new{Performance of the proposed model under varying training data fractions. Results are reported as mean $\pm$ standard deviation over 10 independent runs with fixed validation and test splits throughout all experiments. Training subsets are constructed as consistent nested fractions (50\%–100\%) of the full training set for each run.}}
\resizebox{0.8\linewidth}{!}{
\begin{tabular}{c|c|c|c}
\hline \hline
\textbf{{\diagbox[width=17em]{Training Data Fraction}{Metric}}} & \textbf{RMSE} $\downarrow$ & \textbf{R$^{2}$ $\uparrow$}  & \textbf{PSNR (dB)} $\uparrow$ \\
\hline
0.5 & 0.0361 $\pm$ 0.0107 & 0.8892 $\pm$ 0.0939 & 29.2229 $\pm$ 2.5827 \\
0.6 & 0.0316 $\pm$ 0.0068 & 0.8861 $\pm$ 0.0964 & 30.1900 $\pm$ 1.7241 \\
0.7 & 0.0284 $\pm$ 0.0127 & 0.9153 $\pm$ 0.1412 & 31.6219 $\pm$ 3.2975 \\
0.8 & 0.0263 $\pm$ 0.0066 & 0.9252 $\pm$ 0.0536 & 31.8732 $\pm$ 2.2029 \\
0.9 & 0.0258 $\pm$ 0.0077 & 0.9269 $\pm$ 0.0892 & 32.0994 $\pm$ 2.2910 \\
1 & 0.0227 $\pm$ 0.0050 & 0.9563 $\pm$ 0.0174 & 33.0990 $\pm$ 1.8760
\\ \hline \hline
\end{tabular}
}
\label{tab:abl_train_frac}
\end{table*}

\subsubsection{\new{Sensitivity to Savitzky-Golay Parameters}}
\label{sec:abl_savgol}

\new{To justify the choice of Savitzky-Golay filter parameters, we conduct sensitivity analyses over Savitzky-Golay smoothing filter polynomial order and window size.
As shown in Table~\ref{tab:abl_savgol_d}, a polynomial order of \( d = 2 \) provides the best performance, while lower-order smoothing underfits the spectral structure and higher-order fitting offers no significant improvement.}

\new{Similarly, Table~\ref{tab:abl_savgol_w} shows that window sizes in the range of 9–13 yield optimal performance, with \( W = 11 \) providing a stable and consistent balance between noise suppression and signal preservation. Larger window sizes gradually degrade performance due to over-smoothing effects.}

\new{These results validate the selected hyperparameters and demonstrate that the Savitzky-Golay filter improves prediction accuracy without excessive distortion of spectral features.}

\begin{table*}[]
\centering
\caption{\new{Comparison of Proposed Model with varying Savitzky-Golay smoothing filter polynomial order}}
\resizebox{0.8\linewidth}{!}{
\begin{tabular}{c|c|c|c}
\hline \hline
\textbf{{\diagbox[width=17em]{Polynomial Order}{Metric}}} & \textbf{RMSE} $\downarrow$ & \textbf{R$^{2}$ $\uparrow$}  & \textbf{PSNR (dB)} $\uparrow$ \\
\hline
1 & 0.0264 $\pm$ 0.0051 & 0.8938 $\pm$ 0.0433 & 31.7216 $\pm$ 1.5989 \\
2 & 0.0227 $\pm$ 0.0050 & 0.9563 $\pm$ 0.0174 & 33.0990 $\pm$ 1.8760 \\
4 & 0.0228 $\pm$ 0.0051 & 0.9556 $\pm$ 0.0173 & 33.0508 $\pm$ 1.8875
\\ \hline \hline
\end{tabular}
}
\label{tab:abl_savgol_d}
\end{table*}

\begin{table*}[]
\centering
\caption{\new{Comparison of Proposed Model with varying Savitzky-Golay smoothing filter window size}}
\resizebox{0.8\linewidth}{!}{
\begin{tabular}{c|c|c|c}
\hline \hline
\textbf{{\diagbox[width=14em]{Window Size}{Metric}}} & \textbf{RMSE} $\downarrow$ & \textbf{R$^{2}$ $\uparrow$}  & \textbf{PSNR (dB)} $\uparrow$ \\
\hline
3 & 0.0234 $\pm$ 0.0053 & 0.9514 $\pm$ 0.0183 & 32.8375 $\pm$ 1.9369 \\
5 & 0.0229 $\pm$ 0.0051 & 0.9550 $\pm$ 0.0176 & 33.0167 $\pm$ 1.8943 \\
7 & 0.0228 $\pm$ 0.0051 & 0.9557 $\pm$ 0.0172 & 33.0547 $\pm$ 1.8867 \\
9 & 0.0227 $\pm$ 0.0050 & 0.9561 $\pm$ 0.0173 & 33.0847 $\pm$ 1.8831 \\
11 & 0.0227 $\pm$ 0.0050 & 0.9563 $\pm$ 0.0174 & 33.0990 $\pm$ 1.8760 \\
13 & 0.0227 $\pm$ 0.0050 & 0.9564 $\pm$ 0.0175 & 33.0901 $\pm$ 1.8695 \\
15 & 0.0228 $\pm$ 0.0050 & 0.9563 $\pm$ 0.0177 & 33.0433 $\pm$ 1.8523 \\
17 & 0.0230 $\pm$ 0.0050 & 0.9560 $\pm$ 0.0180 & 32.9653 $\pm$ 1.8212 \\
19 & 0.0232 $\pm$ 0.0049 & 0.9553 $\pm$ 0.0183 & 32.8592 $\pm$ 1.7869 \\
21 & 0.0236 $\pm$ 0.0049 & 0.9543 $\pm$ 0.0188 & 32.7272 $\pm$ 1.7425 \\
23 & 0.0241 $\pm$ 0.0048 & 0.9525 $\pm$ 0.0194 & 32.5390 $\pm$ 1.6855 \\
25 & 0.0248 $\pm$ 0.0047 & 0.9498 $\pm$ 0.0202 & 32.2748 $\pm$ 1.6015
\\ \hline \hline
\end{tabular}
}
\label{tab:abl_savgol_w}
\end{table*}


\subsubsection{\newr{Comparison with Alternative Smoothing Methods}}
\label{sec:abl_smoothing_methods}

\newr{To verify that the performance improvements are specific to the proposed Savitzky--Golay smoothing rather than a generic consequence of post-processing, we compare it with several commonly used one-dimensional smoothing techniques representing different filtering paradigms, including moving average, Gaussian, median, Wiener, Butterworth, and cubic spline smoothing. All smoothing methods are applied only during inference to the same MCSR predictions, while the underlying trained model remains unchanged. The parameters of each smoothing method are selected to provide comparable smoothing strength while following commonly adopted settings in signal processing. Specifically, moving average, median, and Wiener filters use a window length of 11 to match the Savitzky--Golay window.}

\newr{The results are summarized in Table~\ref{tab:abl_smoothing_methods}. All smoothing methods improve the unsmoothed predictions as expected, confirming that suppressing high-frequency prediction noise is generally beneficial for spectral reconstruction. Among all evaluated techniques, Savitzky--Golay smoothing achieves the best overall performance, obtaining the lowest RMSE of \(0.0227 \pm 0.0050\), the highest \(R^2\) score of \(0.9563 \pm 0.0174\), and the highest PSNR of \(33.10 \pm 1.88\) dB. The results demonstrate that the reported improvements are not merely due to generic post-processing. Rather, the Savitzky--Golay filter provides a more favorable balance between high-frequency noise suppression and preservation of local spectral characteristics, making it particularly well suited for post-processing predicted absorption spectra.}

\begin{table*}[]
\centering
\caption{\newr{Comparison of the proposed framework using different spectral smoothing methods applied as post-processing during inference.}}
\resizebox{0.8\linewidth}{!}{
\color{black}
\begin{tabular}{l|l|c|c|c}
\hline \hline
\textbf{Method} & \textbf{Parameters} & \textbf{RMSE} $\downarrow$ & \textbf{R$^{2}$ $\uparrow$}  & \textbf{PSNR (dB)} $\uparrow$ \\
\hline
No Smoothing & -- & 0.0322 $ \pm $ 0.0126 & 0.9031 $ \pm $ 0.0754 & 30.29 $ \pm $ 3.08 \\

Cubic Spline & $s=2$ & 0.0303 $\pm$ 0.0056 & 0.8467 $\pm$ 0.0620 & 30.50 $\pm$ 1.57 \\
Wiener & $W=11$ & 0.0277  $\pm$ 0.0057 & 0.8926 $\pm$ 0.0458 & 31.28 $\pm$ 1.67 \\
Moving Average & $W=11$ & 0.0265 $\pm$ 0.0055 & 0.8937  $\pm$ 0.0457 & 31.68 $\pm$ 1.69 \\
Gaussian & $\sigma=3.0$ & 0.0257 $\pm$ 0.0055 & 0.9049 $\pm$ 0.0420 & 31.98 $\pm$ 1.77 \\
Butterworth & Order $=3$, Cutoff $=0.1$ & 0.0237 $\pm$ 0.0054 & 0.9554 $\pm$ 0.0183 & 32.70 $\pm$ 1.89 \\
Median & $W=11$ & 0.0237 $\pm$ 0.0055 & 0.9380 $\pm$ 0.0293 & 32.72 $\pm$ 1.98 \\

Savitzky--Golay & $W=11,\; d=2$ & 0.0227 $\pm$ 0.0050 & 0.9563 $\pm$ 0.0174 & 33.10 $\pm$ 1.88 \\
\hline \hline
\end{tabular}
}
\label{tab:abl_smoothing_methods}

\vspace{1mm}
\raggedright
\footnotesize
\newr{\textit{Note:} \(W\) denotes the filter window length, \(d\) the polynomial order of the Savitzky--Golay filter, \(\sigma\) the standard deviation of the Gaussian filter, \(s\) the smoothing factor of the cubic spline, and ``Order'' and ``Cutoff'' denote the Butterworth filter order and normalized cutoff frequency, respectively.}

\end{table*}


\subsubsection{\new{Overall Discussion of Ablation Study}}

\new{The comprehensive ablation results provide several important insights into the design of the proposed framework. First, transfer learning emerges as a critical component for effective feature extraction, particularly under this limited data setup. Second, the MCSR module consistently improves spectral representation by capturing localized dependencies, leading to more accurate predictions. Third, Savitzky-Golay smoothing contributes to performance gains by reducing high-frequency noise without significantly distorting the underlying spectral structure, as further supported by the sensitivity analysis.}

\new{The backbone comparison further suggests that lightweight pretrained convolutional architectures such as MobileNetV2 are more suitable in this setting, offering superior generalization and stability. Additionally, the training data analysis confirms that the model maintains competitive performance even with reduced data, indicating that it learns meaningful and generalizable representations rather than relying on memorization.}

\subsection{\new{Impact of Smoothing on Spectral Characteristics}}

\begin{figure}[ht!]
	\centering
	\includegraphics[width=0.7\linewidth]{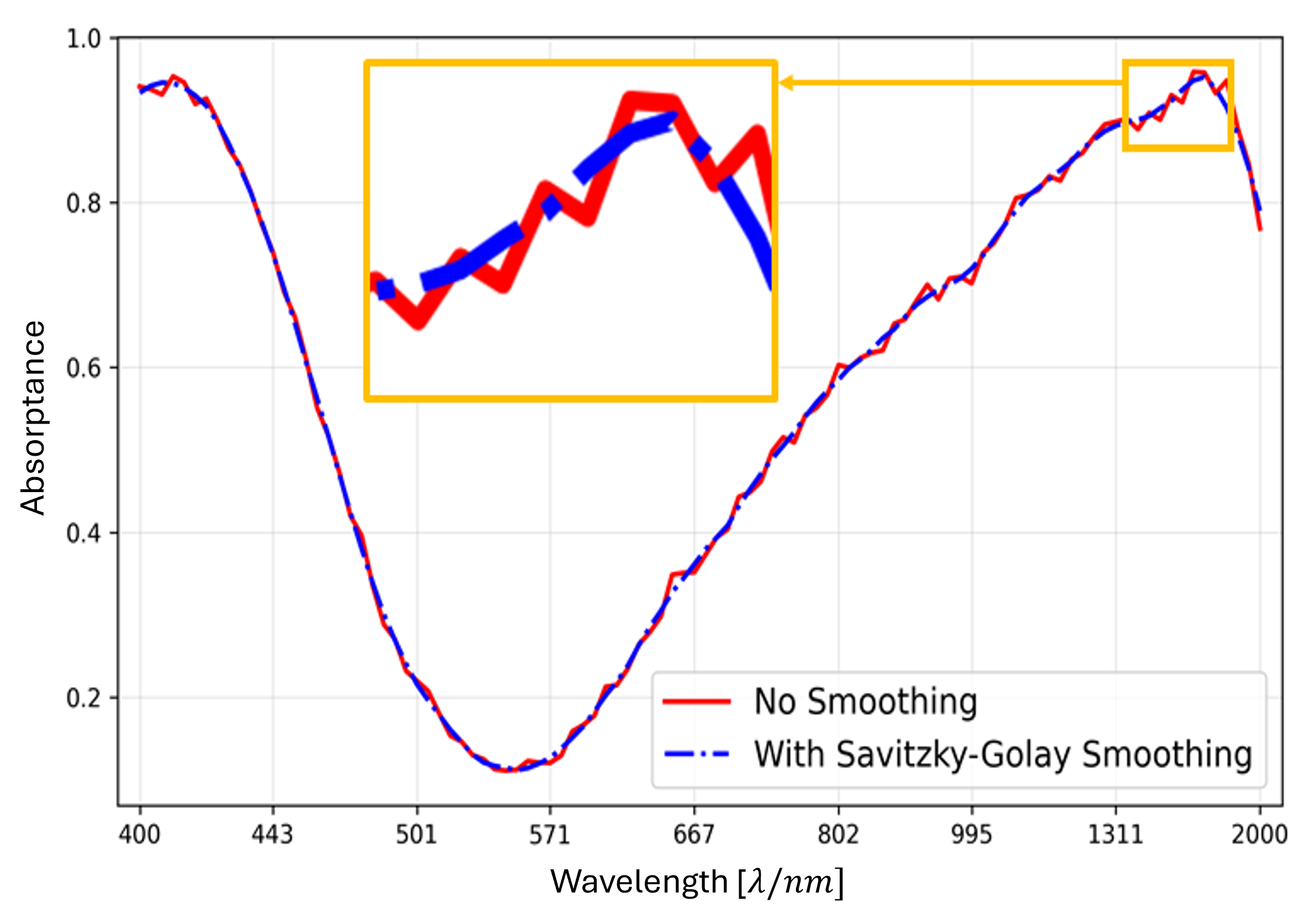}
    \caption{\new{Impact of Savitzky–Golay smoothing on predicted spectra for MXene-based solar absorbers. The smoothed prediction reduces high-frequency fluctuations while preserving the dominant absorption bands, broadband absorption behavior, and peak locations relative to the ground truth. The zoomed regions illustrate improved spectral stability while maintaining physically meaningful spectral characteristics relevant to solar-energy-oriented metasurface design.}}
	\label{fig:savgol_pred}
\end{figure}

\new{To assess the impact of Savitzky--Golay smoothing beyond standard regression metrics and to verify that it does not distort physically meaningful spectral characteristics, we analyze its effect on key spectral features, including peak location, spectral bandwidth, and absorption behavior. Figure~\ref{fig:savgol_pred} compares sample predicted spectra before and after smoothing, showing that Savitzky--Golay smoothing effectively reduces high-frequency fluctuations in the predicted spectra while preserving the overall spectral shape and major absorption features.}

\new{In particular, the positions of dominant absorption peaks remain largely unchanged. Quantitatively, the average peak wavelength shifts from 591.04 nm (unsmoothed) to 593.43 nm after smoothing, compared to 598.19 nm for the ground truth, corresponding to an improvement in peak wavelength error from approximately 7.1 nm to about 4.8 nm. Similarly, the predicted peak amplitude after smoothing (0.8986) remained close to the ground-truth value (0.8939), whereas the unsmoothed prediction exhibited a larger deviation (0.9218). This indicates improved alignment with the ground truth after smoothing. We further analyze the spectral bandwidth using the full-width at half-maximum (FWHM). The average FWHM increases from 352.91 nm (unsmoothed) to 391.33 nm (smoothed), compared to 349.50 nm for the ground truth. This moderate broadening is consistent with the expected effect of smoothing and reflects suppression of high-frequency noise rather than distortion of the underlying resonance behavior.}

\new{Additionally, to assess the impact on application-relevant performance, we evaluate the mean wavelength within the FWHM region. The smoothed predictions yield an average value of 650.42 nm, compared to 643.27 nm for the ground truth, indicating only a small deviation relative to the overall spectral bandwidth. This confirms that the dominant absorption region is preserved. For MXene-based solar absorbers\cite{Waseer2025}, preserving the broadband absorption profile and dominant spectral energy distribution is important for maintaining effective light-harvesting characteristics, indicating that the proposed smoothing strategy does not adversely affect application-relevant absorber behavior.}

\new{Overall, these results demonstrate that Savitzky--Golay smoothing primarily suppresses high-frequency prediction noise without introducing significant distortion of physically relevant spectral features. The observed improvements in RMSE and PSNR therefore reflect enhanced prediction stability rather than artificial error reduction.}

\subsection{Computational Efficiency Analysis}

\begin{figure}[ht!]
	\centering
	\includegraphics[width=0.5\linewidth]{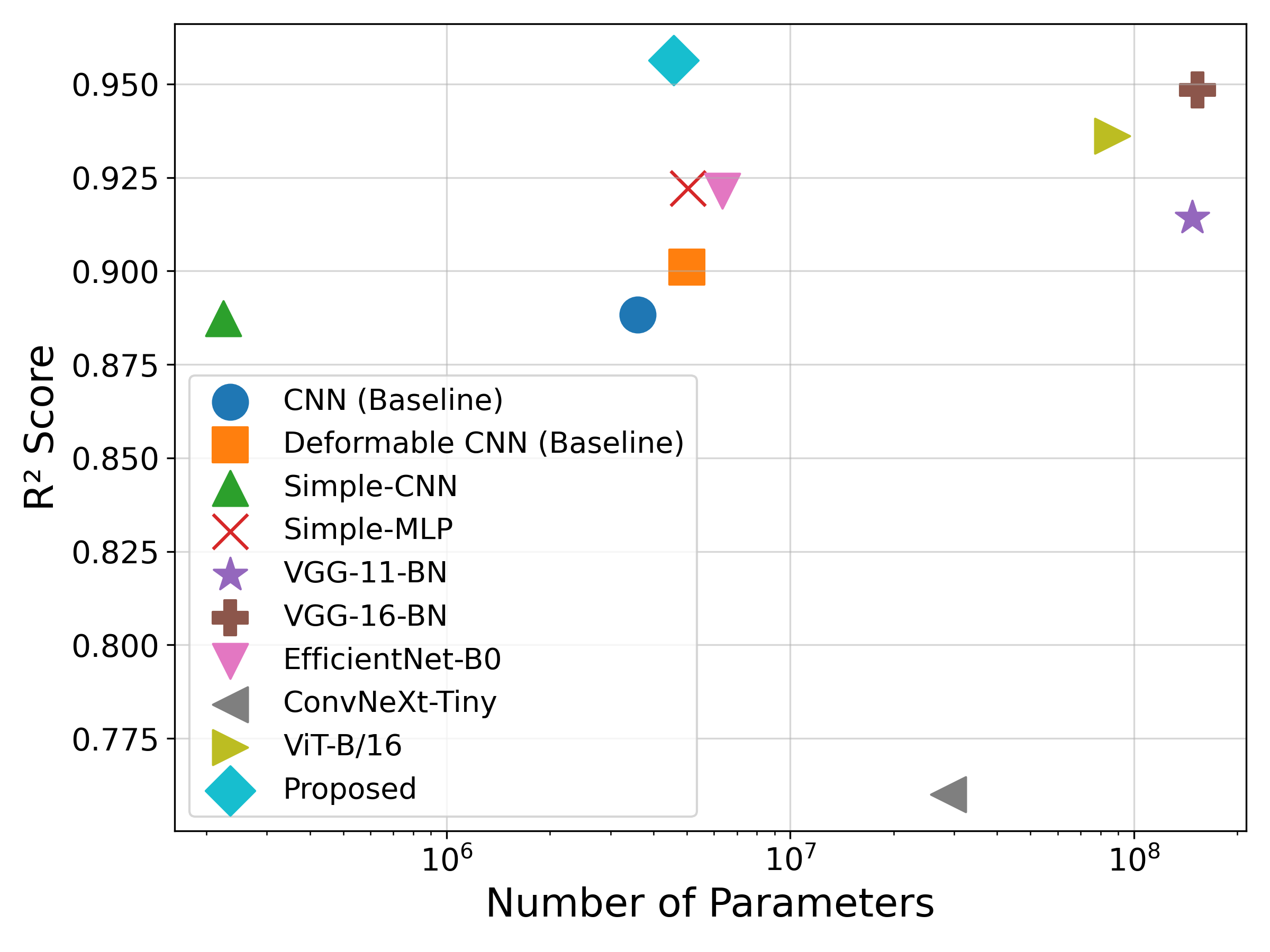}
    \caption{Model complexity (number of parameters, log scale) and \( R^2 \) score for Baseline architectures, and the proposed framework with various backbones.}
	\label{fig:params_vs_r2}
\end{figure}




Figure \ref{fig:params_vs_r2} illustrates the trade-off between the model complexity and predictive performance in terms of \( R^2 \) score for baseline CNN and Deformable CNN architectures \cite{Waseer2025}, as well as several lightweight backbones integrated within the proposed framework. The proposed MobileNetV2-based framework achieves the highest \( R^2 \) score of 0.9563 with a moderate parameter count, outperforming significantly larger backbones such as VGG and ViT; demonstrating the effectiveness of the proposed model in balancing model complexity and predictive performance.

\new{We further analyze the computational complexity of the models in terms of giga floating-point operations (GFLOPs). For an input resolution of $64 \times 64$, the proposed model requires approximately 0.0568 GFLOPs, compared to 0.0506 GFLOPs for the baseline CNN and 0.1582 GFLOPs for the deformable CNN. This indicates that the proposed model achieves higher predictive accuracy with only a modest increase in computational cost relative to the baseline CNN, while remaining significantly more efficient than the deformable CNN.}

\new{The inference time for the proposed model is extremely low, requiring only \(8.49\pm0.08\) milliseconds per sample on CPU and \(3.56\pm0.03\) milliseconds per sample on an NVIDIA A4500 GPU (batch size = 1, mean$\pm$std), enabling near-instantaneous spectral prediction for individual metasurface designs.}
\new{In contrast, CST Microwave Studio simulations rely on the finite-difference time-domain (FDTD) method, which involves iterative time-stepping and spatial discretization of Maxwell’s equations, resulting in substantial computational overhead. Even with available parallelization strategies such as multithreading and GPU acceleration, this iterative process remains computationally expensive. In our setup, simulating a single metasurface design requires approximately 2–3 minutes per sample on a standard workstation (Intel Core i7-7500U CPU with 16 GB RAM).}

\new{This difference reflects fundamentally distinct computational paradigms. The proposed model performs a direct forward mapping from geometry to spectral response, whereas FDTD-based solvers iteratively compute electromagnetic field evolution across space and time. Consequently, the proposed approach enables high-throughput evaluation, where \(10^3\) metasurface designs can be processed within a few seconds, compared to several hours required by conventional simulation tools.}

\new{It is important to emphasize that these approaches serve complementary roles. Full-wave solvers provide detailed electromagnetic field distributions and high-fidelity physical insights, while CNNs including the proposed deep learning framework acts as fast surrogate models for spectral prediction within the learned design space \cite{cheng2024fdtd}. Therefore, the comparison in computational time should be interpreted as indicative of relative efficiency rather than a strict one-to-one benchmark, as a fully controlled comparison would require identical hardware configurations and optimized parallelization settings for both approaches, which are beyond the scope of this work.}

\subsection{Robustness and Explainability via Grad-CAM}

\begin{figure}[!h]
	\centering
	\includegraphics[width=0.9\linewidth]{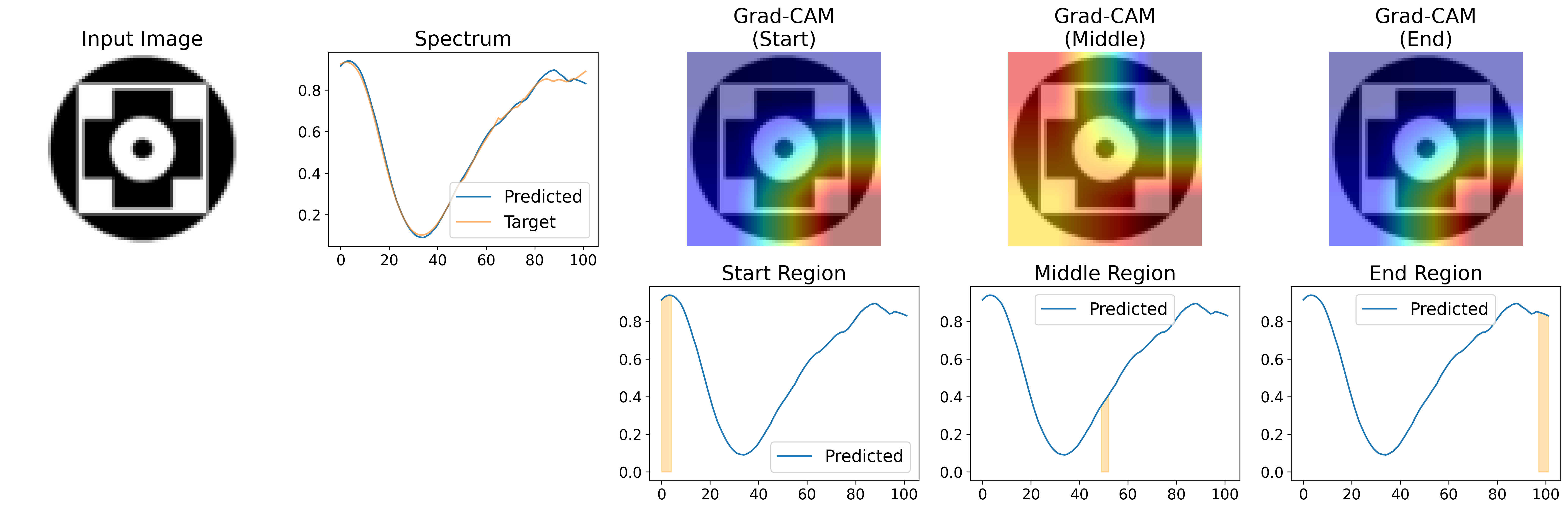}
    \includegraphics[width=0.9\linewidth]{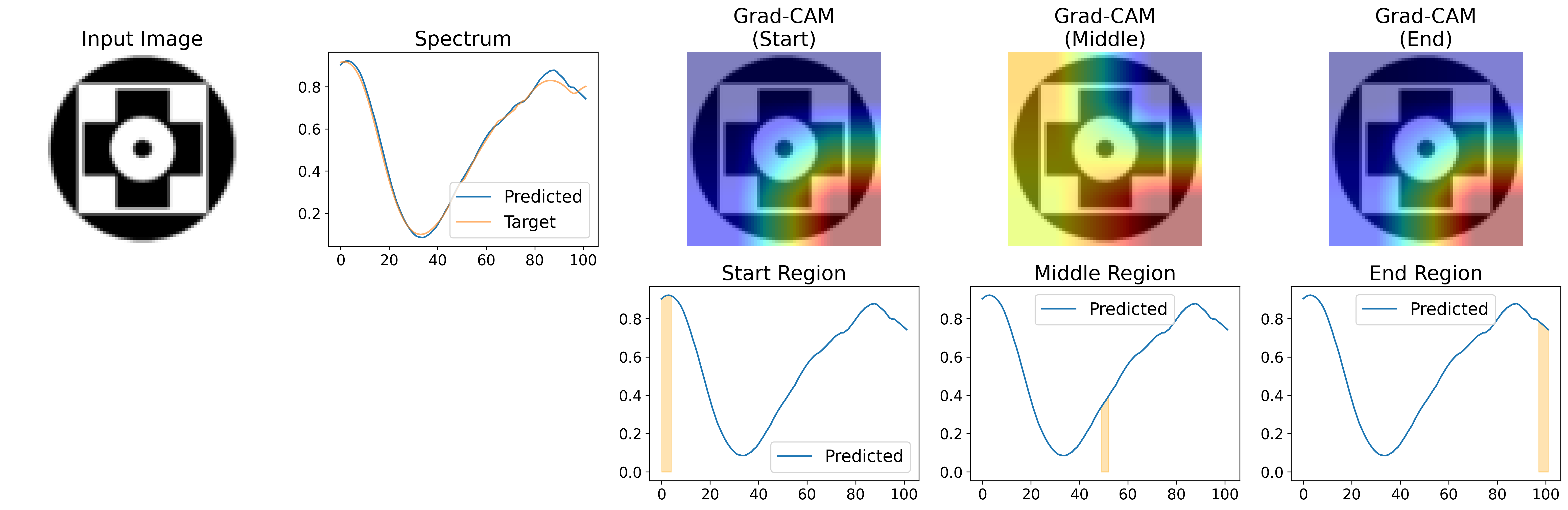}
    \includegraphics[width=0.9\linewidth]{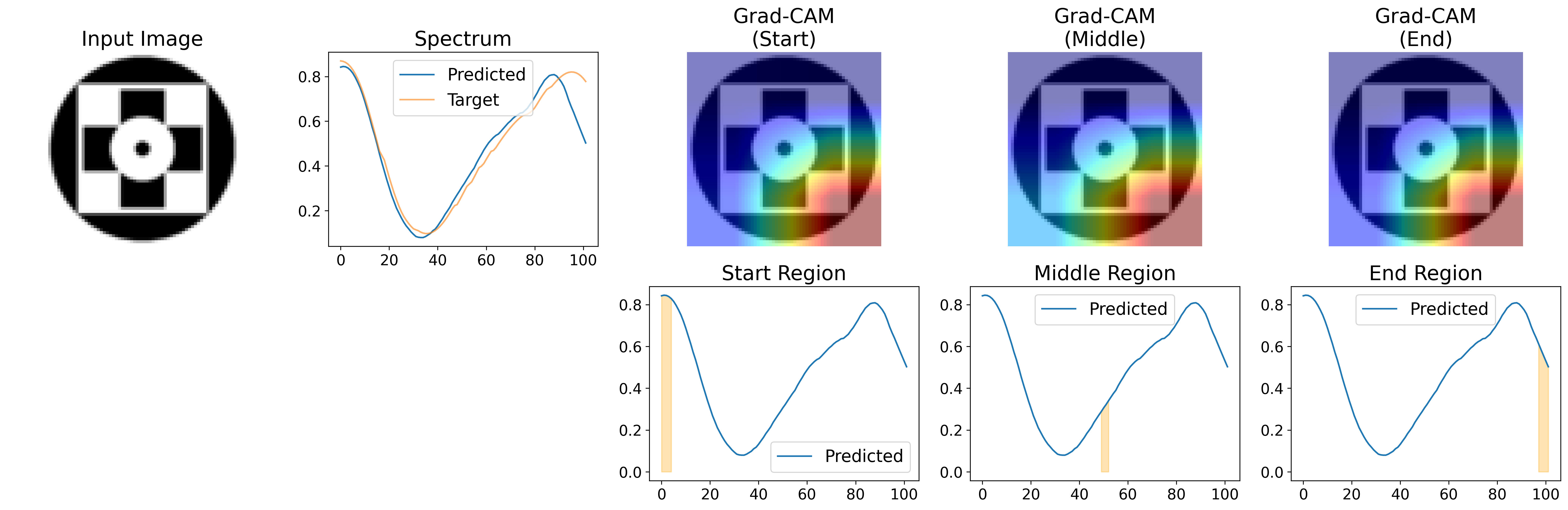}
    \includegraphics[width=0.9\linewidth]{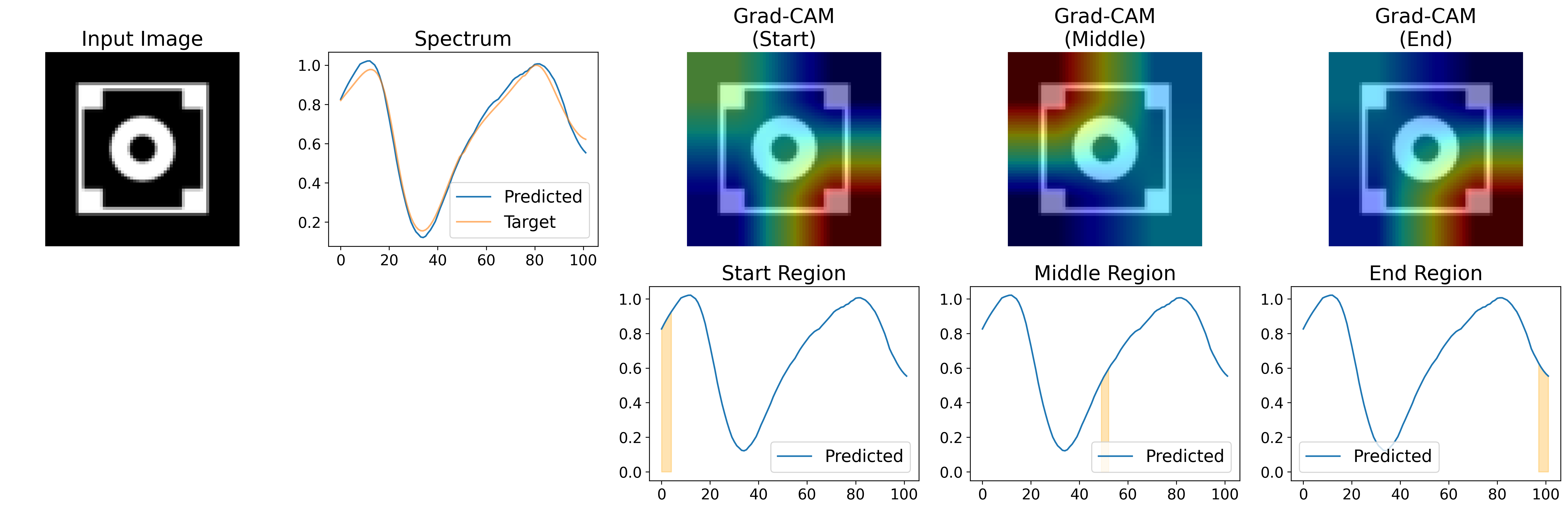}
	\caption{\new{Grad-CAM visualizations illustrating the model’s focus during spectral prediction across four representative metasurface designs. The top row shows the original design, the middle rows display slightly altered designs, and the bottom row presents substantially transformed structures. Despite geometric variations, the model consistently highlights key spectral regions that influence absorption, demonstrating robustness and interpretability. (Best visible in color print)}}
	\label{figGradCAM}
\end{figure}

The robustness and interpretability of the proposed model were further investigated using Grad-CAM visualizations. Grad-CAM provides a visual representation of the regions within the metasurface geometry that predominantly influence spectral predictions. As depicted in Fig.~\ref{figGradCAM}, the model effectively identifies spectral regions that exhibit significant contributions to the absorption spectra, highlighting areas of interest that align with key resonant structures.

The visualizations are structured across four representative metasurface designs, showcasing the model's prediction robustness under varying geometric transformations. The top row in each panel illustrates the original metasurface design, the middle rows present two slightly transformed variants, and the bottom row depicts a substantially altered structure. Despite the geometric variations, the model consistently identifies influential regions, indicating its spatial adaptability and robustness to structural changes.

The Grad-CAM heatmaps reveal that the proposed model not only captures the primary structural features but also emphasizes localized resonant areas that contribute to specific spectral peaks and dips. This ability to visually interpret model predictions enhances its applicability in the design and optimization of MXene-based metasurfaces, enabling researchers to rationalize and refine geometric patterns for targeted spectral responses effectively.

\new{Interestingly, the high-activation Grad-CAM regions frequently coincide with geometrical boundaries, ring edges, corners, and aperture-like regions within the MXene metasurface patterns.} 
\new{This observation suggests that the proposed framework learns physically meaningful spatial features related to resonance formation and absorption behavior rather than relying on arbitrary image patterns.}

Furthermore, the observed alignment between the Grad-CAM heatmaps and known spectral hotspots confirms that the model leverages physically relevant features for spectral prediction, thereby increasing the overall explainability and reliability of the proposed approach.

\subsection{\new{Limitations and Future Work}}

\new{A limitation of the current study is the relatively modest dataset size (500 metasurface designs), which is constrained by the computational cost of full-wave electromagnetic simulations. While transfer learning and repeated evaluation demonstrate strong generalization within the available design space, larger and more diverse datasets could further improve model robustness and enable better generalization to unseen geometries.}

\new{In addition, the current framework is trained exclusively on simulation-generated, noise-free spectra obtained under controlled FDTD conditions. As a result, fabrication imperfections, material variability, and experimental measurement noise are not explicitly modeled in the present study. Furthermore, the proposed framework considers a fixed MXene--SiO$_2$--Ag material stack under normally incident, single-polarization illumination conditions.} \newr{Consequently, the learned model is optimized within this constrained physical design space and may require adaptation when applied to different material systems, multilayer configurations, polarization states, or oblique-incidence conditions. Therefore, the current study should be interpreted as demonstrating adaptation to a representative out-of-distribution physical variation rather than complete domain generalization across arbitrary metasurface configurations. Comprehensive evaluation under multiple unseen physical conditions remains an important direction for future work.}

\new{Another limitation arises from the use of transfer learning from natural-image datasets. Although MobileNetV2 demonstrated strong performance and stability in our experiments, pretrained representations learned from ImageNet may not always be optimal for nanophotonic spectral prediction tasks. Different backbone architectures may exhibit varying trade-offs between representational capacity, robustness, and data efficiency, particularly under limited-data conditions.}

\new{Future work will therefore focus on expanding the dataset diversity, incorporating experimentally measured spectra, exploring alternative backbone architectures and transformer-based models, extending the framework to more complex metasurface configurations, and investigating inverse-design strategies for broader nanophotonic optimization tasks.}

\subsection{\new{Adaptation to New Design Space}}

\begin{figure}[ht!]
	\centering
	\includegraphics[width=0.7\linewidth]{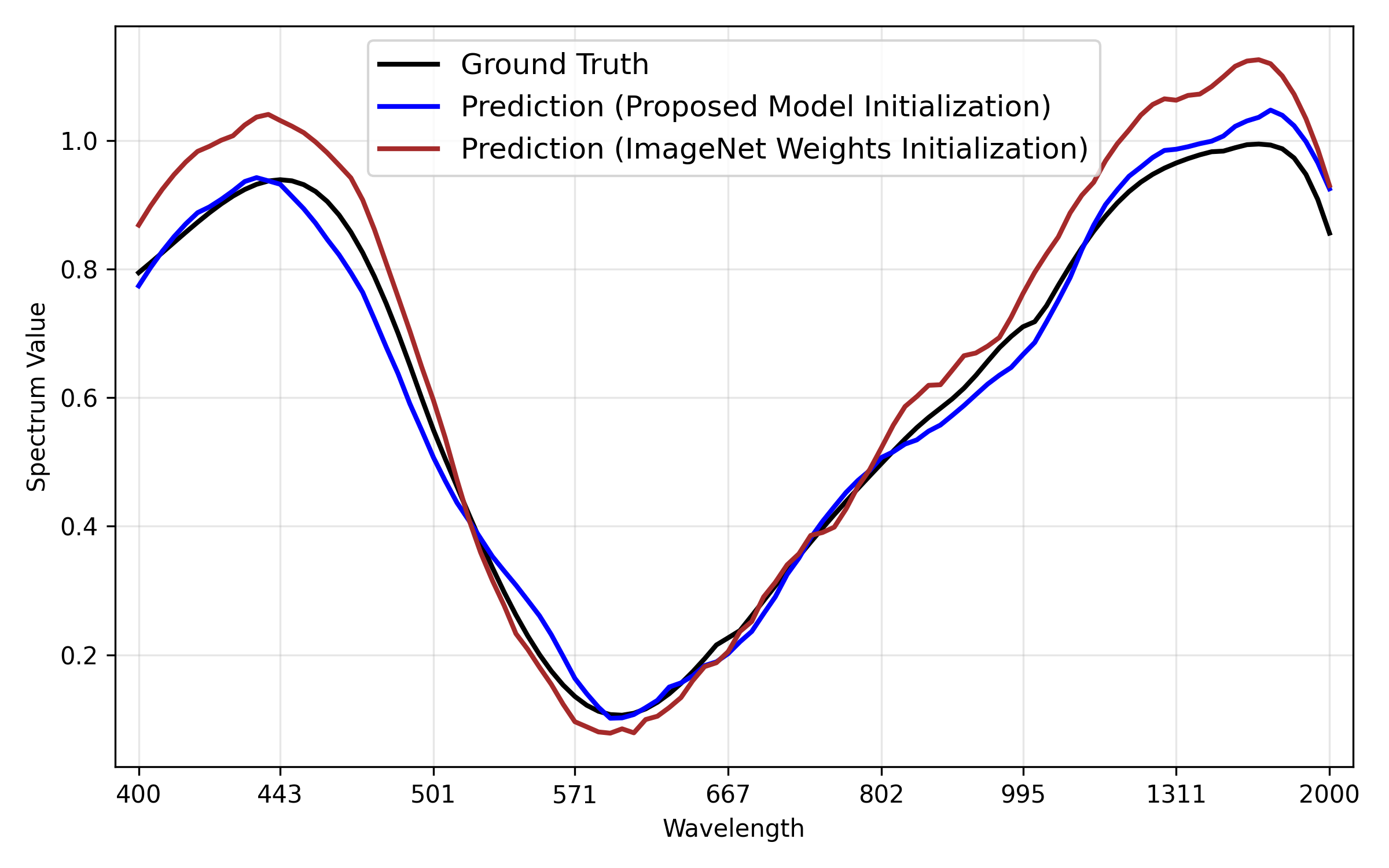}
    \caption{\new{Prediction on the modified design space with increased SiO$_2$ spacer thickness. The proposed-model initialization better follows the ground-truth spectrum than ImageNet-only initialization.}}
	\label{fig:revised_pred}
\end{figure}

\new{To evaluate the generalization capability of the proposed model beyond the training design space, we consider a modified physical configuration in which the SiO$_2$ dielectric spacer thickness is increased from 200 nm to 220 nm, while all other parameters; including the 40 nm MXene top layer, 100 nm Ag reflector, and unit-cell geometries are kept unchanged. This modification shifts the cavity resonance, resulting in spectral responses that are not present in the original training dataset.}

\new{Since the model is trained under fixed physical conditions, it is not expected to directly generalize to this out-of-distribution setting. To assess adaptability, we fine-tune the pretrained model using a small subset of 13 samples from the new configuration and evaluate performance on unseen geometries.}

\new{The experimental results show that the fine-tuned model achieves strong performance, with an average RMSE of 0.0375, \(R^2 = 0.9843\), and PSNR of 28.50 dB. In contrast, a model initialized with standard ImageNet weights achieves significantly lower performance (\(RMSE \approx 0.104, R^2 \approx 0.879\)), indicating that transferring learned representations from the original model provides a substantial advantage. Figure~\ref{fig:revised_pred} illustrates sample predictions, showing that the fine-tuned model closely follows the ground truth spectral response, particularly around shifted resonance regions.}

\new{These results demonstrate that while the proposed framework is optimized for a fixed design space, it can be efficiently adapted to new physical conditions with a small number of additional samples, reinforcing its practical utility for iterative design workflows.} \new{Although only a single out-of-distribution physical variation is considered in this study, it provides an initial validation of the proposed adaptation strategy. Evaluating multiple unseen material systems, layer configurations, and  llumination conditions will be investigated in future work.}

\newr{Although the present experiment considers only a single out-of-distribution configuration (200 nm $\rightarrow$ 220 nm dielectric thickness), it was intentionally selected as a representative case because dielectric thickness directly alters the optical path length and resonance characteristics of the MIM cavity. Consequently, the resulting spectra differ from those observed during training, providing a meaningful evaluation of adaptation to unseen physical conditions. Evaluating multiple material systems, polarization states, incidence angles, and wider geometric variations would require generating entirely new full-wave simulation datasets, which is computationally expensive and therefore left as future work.}

\subsection{\new{Impact of Prediction Errors on Engineering Applications}}

\new{We discuss how deviations in the predicted spectra may affect downstream applications such as solar energy harvesting, thermal emission, and sensing; to assess the engineering implications of prediction errors. For solar harvesting, the impact of spectral prediction error can be estimated by integrating the absolute error with respect to the incident solar spectrum (AM 1.5G) over the wavelength range of 400–2000 nm. Using the observed test-RMSE (0.0227) as a uniform approximation across wavelengths provides an upper bound on the absorbed power error of approximately $\Delta P \lesssim 23$ W/m$^2$.  This estimate is consistent with the relatively low absolute deviations observed across the representative examples shown in Fig.~\ref{fig4}. This level of uncertainty is acceptable for rapid design screening and optimization, although it may be non-negligible for precision harvesting optimization, and more precise estimation may be required for final device design.}

\new{For thermal emission, the considered wavelength range (400–2000 nm) lies outside the peak emission region of typical blackbody sources operating at 300–800 K. Therefore, the predicted spectra are most relevant for near-infrared selective emitters or thermophotovoltaic applications, where moderate spectral deviations are less likely to significantly alter system-level performance.}

\new{For refractive-index sensing applications, prediction errors near resonance dips are more critical. The observed peak absolute error (approximately 0.04) may lead to slight underestimation of dip depth, which could affect the estimation of sensitivity or figure-of-merit (FOM). Accurate sensing analysis would therefore require localized evaluation at resonance wavelengths.}

\section{Conclusion}

\new{The results further demonstrate that the proposed framework derives its effectiveness from the complementary interaction between transfer learning, spectral refinement, and signal-based smoothing.} This study presents a comprehensive deep learning framework for predicting the spectral response of MXene-based metasurface absorbers, integrating transfer learning, multi-channel spectral refinement, and Savitzky-Golay smoothing. The proposed model leverages a pretrained MobileNetV2 architecture, effectively adapted to predict 102-point absorption spectra from $64 \times 64$ grayscale metasurface geometries. Through extensive experimentation, the proposed model demonstrated substantial performance improvements over baseline CNN and deformable CNN architectures, achieving lower RMSE, higher \( R^2 \) scores, and superior PSNR values across all evaluated datasets. 

The inclusion of MCSR and Savitzky-Golay filtering was pivotal in refining spectral outputs and mitigating noise, thereby enhancing spectral prediction accuracy while maintaining a lightweight model architecture comprising only 2.9 million parameters. Furthermore, the Grad-CAM visualizations provided valuable interpretability, revealing the critical spectral regions influencing the absorption response and aligning well with known resonant hotspots within the metasurface structures. 


Future work will focus on expanding the model to handle multi-layered metasurfaces and more complex geometric configurations. Additionally, integrating inverse design methodologies to directly generate metasurface structures for desired spectral responses represents a promising direction for further research. Enhanced interpretability techniques, such as attention mechanisms and layer-wise relevance propagation, will also be explored to provide deeper insights into the model’s decision-making process, further consolidating the proposed framework as a comprehensive solution for nanophotonic design and spectral optimization.



\begin{thebibliography}{35}
\expandafter\ifx\csname natexlab\endcsname\relax\def\natexlab#1{#1}\fi
\providecommand{\url}[1]{\texttt{#1}}
\providecommand{\href}[2]{#2}
\providecommand{\path}[1]{#1}
\providecommand{\DOIprefix}{doi:}
\providecommand{\ArXivprefix}{arXiv:}
\providecommand{\URLprefix}{URL: }
\providecommand{\Pubmedprefix}{pmid:}
\providecommand{\doi}[1]{\href{http://dx.doi.org/#1}{\path{#1}}}
\providecommand{\Pubmed}[1]{\href{pmid:#1}{\path{#1}}}
\providecommand{\bibinfo}[2]{#2}
\ifx\xfnm\relax \def\xfnm[#1]{\unskip,\space#1}\fi
\bibitem[{Tang et~al.(2022)Tang, He, Shi, Liu, Chen, and Dong}]{tang2022topological}
\bibinfo{author}{G.-J. Tang}, \bibinfo{author}{X.-T. He}, \bibinfo{author}{F.-L. Shi}, \bibinfo{author}{J.-W. Liu}, \bibinfo{author}{X.-D. Chen}, \bibinfo{author}{J.-W. Dong},
\newblock \bibinfo{title}{Topological photonic crystals: physics, designs, and applications},
\newblock \bibinfo{journal}{Laser \& Photonics Reviews} \bibinfo{volume}{16} (\bibinfo{year}{2022}) \bibinfo{pages}{2100300}.
\bibitem[{Li et~al.(2021)Li, Miao, Zhang, Wu, Zhang, Du, Han, Sun, and Xu}]{li2021recentplasmonic}
\bibinfo{author}{S.~Li}, \bibinfo{author}{P.~Miao}, \bibinfo{author}{Y.~Zhang}, \bibinfo{author}{J.~Wu}, \bibinfo{author}{B.~Zhang}, \bibinfo{author}{Y.~Du}, \bibinfo{author}{X.~Han}, \bibinfo{author}{J.~Sun}, \bibinfo{author}{P.~Xu},
\newblock \bibinfo{title}{Recent advances in plasmonic nanostructures for enhanced photocatalysis and electrocatalysis},
\newblock \bibinfo{journal}{Advanced Materials} \bibinfo{volume}{33} (\bibinfo{year}{2021}) \bibinfo{pages}{2000086}.
\bibitem[{Cui et~al.(2023)Cui, Al{\`u}, and Pendry}]{cui2023guest}
\bibinfo{author}{T.~J. Cui}, \bibinfo{author}{A.~Al{\`u}}, \bibinfo{author}{J.~B. Pendry},
\newblock \bibinfo{title}{Guest editorial special issue on metamaterials, metadevices, and applications},
\newblock \bibinfo{journal}{IEEE Transactions on Microwave Theory and Techniques} \bibinfo{volume}{71} (\bibinfo{year}{2023}) \bibinfo{pages}{3229--3234}.
\bibitem[{Khorasaninejad and Capasso(2017)}]{khorasaninejad2017metalenses}
\bibinfo{author}{M.~Khorasaninejad}, \bibinfo{author}{F.~Capasso},
\newblock \bibinfo{title}{Metalenses: Versatile multifunctional photonic components},
\newblock \bibinfo{journal}{Science} \bibinfo{volume}{358} (\bibinfo{year}{2017}).
\bibitem[{Huang et~al.(2017)Huang, Song, Reineke, Li, Li, Liu, Zhang, Wang, and Zentgraf}]{huang2017volumetric}
\bibinfo{author}{L.~Huang}, \bibinfo{author}{X.~Song}, \bibinfo{author}{B.~Reineke}, \bibinfo{author}{T.~Li}, \bibinfo{author}{X.~Li}, \bibinfo{author}{J.~Liu}, \bibinfo{author}{S.~Zhang}, \bibinfo{author}{Y.~Wang}, \bibinfo{author}{T.~Zentgraf},
\newblock \bibinfo{title}{Volumetric generation of optical vortices with metasurfaces},
\newblock \bibinfo{journal}{ACS Photonics} \bibinfo{volume}{4} (\bibinfo{year}{2017}) \bibinfo{pages}{338--346}.
\bibitem[{Naveed et~al.(2021)Naveed, Ansari, Kim, Badloe, Kim, Oh, Riaz, Tauqeer, Younis, Saleem et~al.}]{naveed2021optical}
\bibinfo{author}{M.~A. Naveed}, \bibinfo{author}{M.~A. Ansari}, \bibinfo{author}{I.~Kim}, \bibinfo{author}{T.~Badloe}, \bibinfo{author}{J.~Kim}, \bibinfo{author}{D.~K. Oh}, \bibinfo{author}{K.~Riaz}, \bibinfo{author}{T.~Tauqeer}, \bibinfo{author}{U.~Younis}, \bibinfo{author}{M.~Saleem}, et~al.,
\newblock \bibinfo{title}{Optical spin-symmetry breaking for high-efficiency directional helicity-multiplexed metaholograms},
\newblock \bibinfo{journal}{Microsystems \& {N}anoengineering} \bibinfo{volume}{7} (\bibinfo{year}{2021}) \bibinfo{pages}{5}.
\bibitem[{Pfeiffer and Grbic(2014)}]{mehmood2016visible}
\bibinfo{author}{C.~Pfeiffer}, \bibinfo{author}{A.~Grbic},
\newblock \bibinfo{title}{Controlling vector bessel beams with metasurfaces},
\newblock \bibinfo{journal}{Physical Review Applied} \bibinfo{volume}{2} (\bibinfo{year}{2014}) \bibinfo{pages}{044012}.
\bibitem[{Landy et~al.(2008)Landy, Sajuyigbe, Mock, Smith, and Padilla}]{landy2008perfect}
\bibinfo{author}{N.~I. Landy}, \bibinfo{author}{S.~Sajuyigbe}, \bibinfo{author}{J.~J. Mock}, \bibinfo{author}{D.~R. Smith}, \bibinfo{author}{W.~J. Padilla},
\newblock \bibinfo{title}{Perfect metamaterial absorber},
\newblock \bibinfo{journal}{Physical Review Letters} \bibinfo{volume}{100} (\bibinfo{year}{2008}) \bibinfo{pages}{207402}.
\bibitem[{Soni and Misra(2023)}]{soni2023machine}
\bibinfo{author}{M.~Soni}, \bibinfo{author}{S.~Misra},
\newblock \bibinfo{title}{Machine-learning-assisted design of multiband terahertz metamaterial absorber},
\newblock \bibinfo{journal}{ACS Applied Optical Materials} \bibinfo{volume}{1} (\bibinfo{year}{2023}) \bibinfo{pages}{1679--1687}.
\bibitem[{Ding et~al.(2023)Ding, Su, Hakimi, Luo, Li, Zhou, Ye, and Yao}]{ding2023machine}
\bibinfo{author}{Z.~Ding}, \bibinfo{author}{W.~Su}, \bibinfo{author}{F.~Hakimi}, \bibinfo{author}{Y.~Luo}, \bibinfo{author}{W.~Li}, \bibinfo{author}{Y.~Zhou}, \bibinfo{author}{L.~Ye}, \bibinfo{author}{H.~Yao},
\newblock \bibinfo{title}{Machine learning in prediction of mxenes-based metasurface absorber for maximizing solar spectral absorption},
\newblock \bibinfo{journal}{Solar Energy Materials and Solar Cells} \bibinfo{volume}{262} (\bibinfo{year}{2023}) \bibinfo{pages}{112563}.
\bibitem[{Mizuno et~al.(2009)Mizuno, Ishii, Kishida, Hayamizu, Yasuda, Futaba, Yumura, and Hata}]{mizuno2009black}
\bibinfo{author}{K.~Mizuno}, \bibinfo{author}{J.~Ishii}, \bibinfo{author}{H.~Kishida}, \bibinfo{author}{Y.~Hayamizu}, \bibinfo{author}{S.~Yasuda}, \bibinfo{author}{D.~N. Futaba}, \bibinfo{author}{M.~Yumura}, \bibinfo{author}{K.~Hata},
\newblock \bibinfo{title}{A black body absorber from vertically aligned single-walled carbon nanotubes},
\newblock \bibinfo{journal}{Proceedings of the National Academy of Sciences} \bibinfo{volume}{106} (\bibinfo{year}{2009}) \bibinfo{pages}{6044--6047}.
\bibitem[{Chen et~al.(2022)Chen, Deng, Zou, Zhao, Liu, Wang, He, Gao, Zhao, and Li}]{chen2022plasmonic}
\bibinfo{author}{K.~Chen}, \bibinfo{author}{C.~Deng}, \bibinfo{author}{C.~Zou}, \bibinfo{author}{Z.~Zhao}, \bibinfo{author}{Q.~Liu}, \bibinfo{author}{X.~Wang}, \bibinfo{author}{L.~He}, \bibinfo{author}{F.~Gao}, \bibinfo{author}{W.~Zhao}, \bibinfo{author}{S.~Li},
\newblock \bibinfo{title}{Plasmonic hot-hole injection combined with patterned substrate for performance improvement in trapezoidal pin gan microwire self-powered ultraviolet photodetector},
\newblock \bibinfo{journal}{Nano Energy} \bibinfo{volume}{104} (\bibinfo{year}{2022}) \bibinfo{pages}{107926}.
\bibitem[{Yao et~al.(2012)Yao, Jin, and Krein}]{yao2012highly}
\bibinfo{author}{W.~Yao}, \bibinfo{author}{J.-M. Jin}, \bibinfo{author}{P.~T. Krein},
\newblock \bibinfo{title}{A highly efficient domain decomposition method applied to 3-d finite-element analysis of electromechanical and electric machine problems},
\newblock \bibinfo{journal}{IEEE Transactions on Energy Conversion} \bibinfo{volume}{27} (\bibinfo{year}{2012}) \bibinfo{pages}{1078--1086}.
\bibitem[{Moharam et~al.(1995)Moharam, Pommet, Grann, and Gaylord}]{moharam1995stable}
\bibinfo{author}{M.~Moharam}, \bibinfo{author}{D.~A. Pommet}, \bibinfo{author}{E.~B. Grann}, \bibinfo{author}{T.~K. Gaylord},
\newblock \bibinfo{title}{Stable implementation of the rigorous coupled-wave analysis for surface-relief gratings: enhanced transmittance matrix approach},
\newblock \bibinfo{journal}{JOSA A} \bibinfo{volume}{12} (\bibinfo{year}{1995}) \bibinfo{pages}{1077--1086}.
\bibitem[{Liu et~al.(2018)Liu, Tan, Khoram, and Yu}]{liu2018training}
\bibinfo{author}{D.~Liu}, \bibinfo{author}{Y.~Tan}, \bibinfo{author}{E.~Khoram}, \bibinfo{author}{Z.~Yu},
\newblock \bibinfo{title}{Training deep neural networks for the inverse design of nanophotonic structures},
\newblock \bibinfo{journal}{Acs Photonics} \bibinfo{volume}{5} (\bibinfo{year}{2018}) \bibinfo{pages}{1365--1369}.
\bibitem[{Park et~al.(2021)Park, Lee, Khan, Wahab, and Kim}]{park2021sersnet}
\bibinfo{author}{S.~Park}, \bibinfo{author}{J.~Lee}, \bibinfo{author}{S.~Khan}, \bibinfo{author}{A.~Wahab}, \bibinfo{author}{M.~Kim},
\newblock \bibinfo{title}{Sersnet: Surface-enhanced raman spectroscopy based biomolecule detection using deep neural network},
\newblock \bibinfo{journal}{Biosensors} \bibinfo{volume}{11} (\bibinfo{year}{2021}) \bibinfo{pages}{490}.
\bibitem[{Khan et~al.(2021)Khan, Huh, and Ye}]{khan2021switchable}
\bibinfo{author}{S.~Khan}, \bibinfo{author}{J.~Huh}, \bibinfo{author}{J.~C. Ye},
\newblock \bibinfo{title}{Switchable and tunable deep beamformer using adaptive instance normalization for medical ultrasound},
\newblock \bibinfo{journal}{IEEE Transactions on Medical Imaging} \bibinfo{volume}{41} (\bibinfo{year}{2021}) \bibinfo{pages}{266--278}.
\bibitem[{Park et~al.(2023)Park, Wahab, Kim, and Khan}]{park2023self}
\bibinfo{author}{S.~Park}, \bibinfo{author}{A.~Wahab}, \bibinfo{author}{M.~Kim}, \bibinfo{author}{S.~Khan},
\newblock \bibinfo{title}{Self-supervised learning for inter-laboratory variation minimization in surface-enhanced raman scattering spectroscopy},
\newblock \bibinfo{journal}{Analyst} \bibinfo{volume}{148} (\bibinfo{year}{2023}) \bibinfo{pages}{1473--1482}.
\bibitem[{Waseer et~al.(2025)Waseer, Baqir, Saqlain, Mughal, and Khan}]{Waseer2025}
\bibinfo{author}{W.~I. Waseer}, \bibinfo{author}{M.~A. Baqir}, \bibinfo{author}{M.~Saqlain}, \bibinfo{author}{M.~J. Mughal}, \bibinfo{author}{S.~Khan},
\newblock \bibinfo{title}{Predictive modeling of mxene-based solar absorbers using a deep neural network},
\newblock \bibinfo{journal}{Journal of the Optical Society of America B} \bibinfo{volume}{42} (\bibinfo{year}{2025}) \bibinfo{pages}{763--772}.
\bibitem[{Dai et~al.(2017)Dai, Qi, Xiong, Li, Zhang, Hu, and Wei}]{dai2017deformable}
\bibinfo{author}{J.~Dai}, \bibinfo{author}{H.~Qi}, \bibinfo{author}{Y.~Xiong}, \bibinfo{author}{Y.~Li}, \bibinfo{author}{G.~Zhang}, \bibinfo{author}{H.~Hu}, \bibinfo{author}{Y.~Wei},
\newblock \bibinfo{title}{Deformable convolutional networks},
\newblock in: \bibinfo{booktitle}{Proceedings of the IEEE international conference on computer vision}, \bibinfo{year}{2017}, pp. \bibinfo{pages}{764--773}.
\bibitem[{Zhu et~al.(2019)Zhu, Hu, Lin, and Dai}]{zhu2019deformable}
\bibinfo{author}{X.~Zhu}, \bibinfo{author}{H.~Hu}, \bibinfo{author}{S.~Lin}, \bibinfo{author}{J.~Dai},
\newblock \bibinfo{title}{Deformable convnets v2: More deformable, better results},
\newblock in: \bibinfo{booktitle}{Proceedings of the IEEE/CVF Conference on Computer Vision and Pattern Recognition}, \bibinfo{year}{2019}, pp. \bibinfo{pages}{9308--9316}.
\bibitem[{Howard et~al.(2019)Howard, Sandler, and Chu}]{howard2019searching}
\bibinfo{author}{A.~Howard}, \bibinfo{author}{M.~Sandler}, \bibinfo{author}{G.~Chu},
\newblock \bibinfo{title}{Searching for mobilenetv3},
\newblock \bibinfo{journal}{Proceedings of the IEEE/CVF International Conference on Computer Vision}  (\bibinfo{year}{2019}) \bibinfo{pages}{1314--1324}.
\bibitem[{Sandler et~al.(2018)Sandler, Howard, Zhu, Zhmoginov, and Chen}]{sandler2018mobilenetv2}
\bibinfo{author}{M.~Sandler}, \bibinfo{author}{A.~Howard}, \bibinfo{author}{M.~Zhu}, \bibinfo{author}{A.~Zhmoginov}, \bibinfo{author}{L.-C. Chen},
\newblock \bibinfo{title}{Mobilenetv2: Inverted residuals and linear bottlenecks},
\newblock in: \bibinfo{booktitle}{Proceedings of the IEEE conference on computer vision and pattern recognition}, \bibinfo{year}{2018}, pp. \bibinfo{pages}{4510--4520}.
\bibitem[{Savitzky and Golay(1964)}]{savitzky1964smoothing}
\bibinfo{author}{A.~Savitzky}, \bibinfo{author}{M.~J.~E. Golay},
\newblock \bibinfo{title}{Smoothing and differentiation of data by simplified least squares procedures},
\newblock \bibinfo{journal}{Analytical Chemistry} \bibinfo{volume}{36} (\bibinfo{year}{1964}) \bibinfo{pages}{1627--1639}.
\bibitem[{Chaudhuri et~al.(2018)Chaudhuri, Alhabeb, Wang, Shalaev, Gogotsi, and Boltasseva}]{chaudhuri2018highly}
\bibinfo{author}{K.~Chaudhuri}, \bibinfo{author}{M.~Alhabeb}, \bibinfo{author}{Z.~Wang}, \bibinfo{author}{V.~M. Shalaev}, \bibinfo{author}{Y.~Gogotsi}, \bibinfo{author}{A.~Boltasseva},
\newblock \bibinfo{title}{Highly broadband absorber using plasmonic titanium carbide (mxene)},
\newblock \bibinfo{journal}{Acs Photonics} \bibinfo{volume}{5} (\bibinfo{year}{2018}) \bibinfo{pages}{1115--1122}.
\bibitem[{Mauchamp et~al.(2014)Mauchamp, Bugnet, Bellido, Botton, Moreau, Magne, Naguib, Cabioc'h, and Barsoum}]{mauchamp2014enhanced}
\bibinfo{author}{V.~Mauchamp}, \bibinfo{author}{M.~Bugnet}, \bibinfo{author}{E.~P. Bellido}, \bibinfo{author}{G.~A. Botton}, \bibinfo{author}{P.~Moreau}, \bibinfo{author}{D.~Magne}, \bibinfo{author}{M.~Naguib}, \bibinfo{author}{T.~Cabioc'h}, \bibinfo{author}{M.~W. Barsoum},
\newblock \bibinfo{title}{Enhanced and tunable surface plasmons in two-dimensional ti 3 c 2 stacks: Electronic structure versus boundary effects},
\newblock \bibinfo{journal}{Physical Review B} \bibinfo{volume}{89} (\bibinfo{year}{2014}) \bibinfo{pages}{235428}.
\bibitem[{Kingma(2014)}]{kingma2014adam}
\bibinfo{author}{D.~P. Kingma},
\newblock \bibinfo{title}{Adam: A method for stochastic optimization},
\newblock \bibinfo{journal}{arXiv preprint arXiv:1412.6980}  (\bibinfo{year}{2014}).
\bibitem[{He et~al.(2015)He, Zhang, Ren, and Sun}]{he2015delving}
\bibinfo{author}{K.~He}, \bibinfo{author}{X.~Zhang}, \bibinfo{author}{S.~Ren}, \bibinfo{author}{J.~Sun},
\newblock \bibinfo{title}{Delving deep into rectifiers: Surpassing human-level performance on imagenet classification},
\newblock in: \bibinfo{booktitle}{Proceedings of the IEEE international conference on computer vision}, \bibinfo{year}{2015}, pp. \bibinfo{pages}{1026--1034}.
\bibitem[{Glorot and Bengio(2010)}]{glorot2010understanding}
\bibinfo{author}{X.~Glorot}, \bibinfo{author}{Y.~Bengio},
\newblock \bibinfo{title}{Understanding the difficulty of training deep feedforward neural networks},
\newblock in: \bibinfo{booktitle}{Proceedings of the thirteenth international conference on artificial intelligence and statistics}, \bibinfo{organization}{JMLR Workshop and Conference Proceedings}, \bibinfo{year}{2010}, pp. \bibinfo{pages}{249--256}.
\bibitem[{Selvaraju et~al.(2017)Selvaraju, Cogswell, Das, Vedantam, Parikh, and Batra}]{selvaraju2017grad}
\bibinfo{author}{R.~R. Selvaraju}, \bibinfo{author}{M.~Cogswell}, \bibinfo{author}{A.~Das}, \bibinfo{author}{R.~Vedantam}, \bibinfo{author}{D.~Parikh}, \bibinfo{author}{D.~Batra},
\newblock \bibinfo{title}{Grad-cam: Visual explanations from deep networks via gradient-based localization},
\newblock in: \bibinfo{booktitle}{Proceedings of the IEEE International Conference on Computer Vision (ICCV)}, \bibinfo{year}{2017}, pp. \bibinfo{pages}{618--626}.
\bibitem[{Simonyan and Zisserman(2014)}]{simonyan2014very}
\bibinfo{author}{K.~Simonyan}, \bibinfo{author}{A.~Zisserman},
\newblock \bibinfo{title}{Very deep convolutional networks for large-scale image recognition},
\newblock \bibinfo{journal}{arXiv preprint arXiv:1409.1556}  (\bibinfo{year}{2014}).
\bibitem[{Tan and Le(2019)}]{tan2019efficientnet}
\bibinfo{author}{M.~Tan}, \bibinfo{author}{Q.~Le},
\newblock \bibinfo{title}{Efficientnet: Rethinking model scaling for convolutional neural networks},
\newblock in: \bibinfo{booktitle}{International conference on machine learning}, \bibinfo{organization}{PMLR}, \bibinfo{year}{2019}, pp. \bibinfo{pages}{6105--6114}.
\bibitem[{Liu et~al.(2022)Liu, Mao, Wu, Feichtenhofer, Darrell, and Xie}]{liu2022convnet}
\bibinfo{author}{Z.~Liu}, \bibinfo{author}{H.~Mao}, \bibinfo{author}{C.-Y. Wu}, \bibinfo{author}{C.~Feichtenhofer}, \bibinfo{author}{T.~Darrell}, \bibinfo{author}{S.~Xie},
\newblock \bibinfo{title}{A convnet for the 2020s},
\newblock in: \bibinfo{booktitle}{Proceedings of the IEEE/CVF conference on computer vision and pattern recognition}, \bibinfo{year}{2022}, pp. \bibinfo{pages}{11976--11986}.
\bibitem[{Dosovitskiy et~al.(2020)Dosovitskiy, Beyer, Kolesnikov, Weissenborn, Zhai, Unterthiner, Dehghani, Minderer, Heigold, Gelly et~al.}]{dosovitskiy2020image}
\bibinfo{author}{A.~Dosovitskiy}, \bibinfo{author}{L.~Beyer}, \bibinfo{author}{A.~Kolesnikov}, \bibinfo{author}{D.~Weissenborn}, \bibinfo{author}{X.~Zhai}, \bibinfo{author}{T.~Unterthiner}, \bibinfo{author}{M.~Dehghani}, \bibinfo{author}{M.~Minderer}, \bibinfo{author}{G.~Heigold}, \bibinfo{author}{S.~Gelly}, et~al.,
\newblock \bibinfo{title}{An image is worth 16x16 words: Transformers for image recognition at scale},
\newblock \bibinfo{journal}{arXiv preprint arXiv:2010.11929}  (\bibinfo{year}{2020}).
\bibitem[{Cheng et~al.(2024)Cheng, Huang, Zhang, Yang, and Zhang}]{cheng2024fdtd}
\bibinfo{author}{Y.~Cheng}, \bibinfo{author}{S.~Huang}, \bibinfo{author}{X.~Zhang}, \bibinfo{author}{S.~Yang}, \bibinfo{author}{X.~Zhang},
\newblock \bibinfo{title}{Fdtd-equivalent neural network model for electromagnetic simulations},
\newblock in: \bibinfo{booktitle}{2024 IEEE International Symposium on Antennas and Propagation and INC/USNC-URSI Radio Science Meeting (AP-S/INC-USNC-URSI)}, \bibinfo{organization}{IEEE}, \bibinfo{year}{2024}, pp. \bibinfo{pages}{2585--2586}.

\end{thebibliography}
\end{document}